%% file: main.tex
\documentclass[twocolumn,twocolappendix]{./aastex631/aastex631}
\shorttitle{Ultrastripped Envelope SN 2021agco}
\shortauthors{Yan et al.}

\usepackage{graphicx}	
\usepackage{amsmath}	
\usepackage{amssymb}	
\usepackage{textcomp}
\usepackage{gensymb}
\usepackage{subfigure}
\usepackage{hyperref}
\usepackage{threeparttable}
\usepackage{ulem}
\usepackage{hyperref}
\usepackage[version=4]{mhchem}
\usepackage{longtable}      
\usepackage{multirow}
\graphicspath{{./}{figures/}}

\newcommand{\msun}{$\mathrm{M_{\odot}}$}
\newcommand{\rsun}{$\mathrm{R_{\odot}}$}
\newcommand{\zsun}{$\mathrm{Z_{\odot}}$}
\newcommand{\Ni}{$^{56}$Ni\ }

\newcommand{\Ha}{H$\alpha$}
\newcommand{\kms}{$\mathrm{km}~\mathrm{s}^{-1}$}

\received{XXX XX, XXXX}
\revised{XXX XX, XXXX}
\accepted{\today}
\submitjournal{ApJ}

\begin{document}

\title{Discovery of the Closest Ultrastripped Supernova: SN 2021agco in UGC 3855}

\input{author_name.tex}

\begin{abstract}
We present the discovery and studies of the helium-rich, fast-evolving supernova (SN) 2021agco at a distance of $\sim$ 40 Mpc. Its early-time flux is found to rise from half peak to the peak of $-16.06\pm0.42$ mag in the $r$ band within $2.4^{+1.5}_{-1.0}$ days, and the post-peak light curves also decline at a much faster pace relative to normal stripped-envelope SNe of Type Ib/Ic. The early-time spectrum of SN~2021agco ($t \approx 1.0$ days after the peak) is characterized by a featureless blue continuum superimposed with a weak emission line of ionized \ion{C}{3}, and the subsequent spectra show prominent \ion{He}{1} lines. Both the photometric and spectroscopic evolution shows close resemblances to SN 2019dge, which is believed to have an extremely stripped progenitor. We reproduce the multicolor light curves of SN 2021agco with a model combining shock-cooling emission with \Ni decay. The best-fit results give an ejecta mass of $\approx 0.3$~\msun\ and a synthesized nickel mass of $\approx 2.2\times10^{-2}$~\msun. The progenitor is estimated to have an envelope radius $R_{\rm env} \approx 80$~\rsun\ and a mass $M_{\rm env} \approx 0.10$~\msun. All these suggest that SN~2021agco can be categorized as an ultrastripped SN~Ib, representing the closest object of this rare subtype. This SN is found to explode in the disk of an Sab-type galaxy with an age of $\sim 10.0$~Gyr and low star-forming activity. Compared to normal SNe Ib/c, the host galaxies of SN 2021agco and other ultrastripped SNe tend to have relatively lower metallicity, which complicates the properties of their progenitor populations. 

\end{abstract}
\section{Introduction}

Massive stars with initial masses $> 8$--10~\msun\ usually end their lives as core-collapse explosions and finally appear as Type II or Type Ibc supernovae (SNe). SNe~II show prominent hydrogen features in optical spectra, suggesting that presupernova stars keep most or at least part of the hydrogen envelopes before explosion.

In contrast, the spectra of SNe~Ib and SNe~Ic lack hydrogen and even helium features (respectively), indicating that the progenitor stars have lost almost all of their H and even He envelopes before exploding \citep[e.g.,][]{optical_spec_SNe_Filippenko1997,2017hsn..book..195G}. Thus, SNe~Ib/Ic are also called as stripped-envelope SNe (SESNe), and their progenitors are believed to have properties similar to Wolf-Rayet stars which could form from single massive stars or binary systems \citep{Smith_N_2011MNRAS}. Currently, there are only a few SNe~Ib/Ic with direct progenitor identifications \citep{Xiang_2017ein_2019ApJ,2018MNRAS.480.2072K,2021MNRAS.504.2073K}, perhaps owing to their progenitors being relatively faint \citep{Yoon_2010ApJ}.

In recent years, efforts of transient surveys have led to discoveries of many weird stellar explosions and shown that the death of massive stars has diverse landscapes. One particularly interesting subclass is the one called ``fast-evolving blue optical transient'' \citep[FBOT;][]{Drout_2014ApJ}, whose members are usually very blue, luminous, and exhibit extremely fast evolution. The well-known example is AT~2018cow, which has a peak luminosity of $M_V \approx -21.0$~mag and a rise time of $< 3.0$ days \citep{AT2018cow_Prentice_2018, AT2018cow_Perley_2019, Xiang_2018cow_2021ApJ}. Following the discovery of AT~2018cow, several similar events were found, including AT~2018lug \citep{ZTF18abvkwla_Ho_2020}, AT~2020xnd \citep{Perley_20xnd_2021MNRAS}, and AT~2020mrf \citep{Yao_2020mrf_2022ApJ}. Nevertheless, the nature of AT~2018cow-like events remains debated, although an origin in massive stars is favored because of their connections with star-forming environments \citep{AT2018cow_Perley_2019, Xiang_2018cow_2021ApJ}. The fast evolution and high luminosity can be interpreted as interaction with the circumstellar material (CSM) surrounding the exploding star. 

In addition to the AT~2018cow-like FBOTs, there is another subclass of subluminous, fast-evolving events that has recently started to populate the atlas of fast transients. They show spectral features similar to those of SNe~Ib/Ic but are relatively faint, with a peak luminosity $> -17.0$ mag and a typical rise time $< 2$--3 days. As the progenitor envelope has been extremely stripped before explosion, they are dubbed ultrastripped-envelope SNe \citep[USSNe;][]{tauris_USSN_2013, tauris_USSN_2015, tauris_USSN_2017,suwa_nu_driven_2015}. Currently known USSNe are very rare, including only iPTF14gqr \citep{iPTF14gqr_De_2018}, SN~2019dge \citep{SN2019dge_2020ApJ}, and SN~2019wxt \citep{AT2019wxt_2022arxiv,SN2019wxt_Agudo}. The object iPTF14gqr was classified as an USSN of Type Ic, while SN 2019dge and SN~2019wxt was classified as an USSNe of Type Ib. Although these USSNe share many similarities with some Ca-rich SNe such as iPTF16hgs \citep{iPTF_16hgs_De_2018ApJ,Moriya_16hgs_2017MNRAS} and SN~2019ehk\citep{SN2019ehk_Jacobson_2020ApJ, SN2019ehk_Nakaoka_2021ApJ}, but they occurred preferentially in star forming environments and may have different progenitor origins.

All the above three USSNe are found to exhibit a fast rise in their early-time light curves, which is explained by shock-cooling mission \citep{SCE_Nakar_2014, SCE_Piro_2015, SCE_Piro_2021}, and the subsequent light curves are powered by radioactive decay of \Ni. To reproduce the light curves of fast-faint events, scenarios with a low ejecta mass or large amounts of fallback ejecta have been proposed \citep{tauris_USSN_2013, tauris_USSN_2015, tauris_USSN_2017, suwa_nu_driven_2015}. These involve a close binary system with a neutron star (NS), which can significantly strip the envelope mass of the surviving massive star \citep{Yoon_2010ApJ}. Its explosion as an USSN would finally lead to the formation of a close NS binary \citep[BNS;][]{tauris_USSN_2015}. The violent merger of binary NS systems could generate strong gravitational waves and the accompanying electromagnetic radiation \citep[i.e., kilonova;][]{GW170817_Abbott_2017}. 

In this paper, we present the discovery and study of a new ultrastripped supernova, SN~2021agco. Section \ref{sec: Observation_properties} describes the observations and the properties of light curves and spectra. In Section \ref{sec: modeling of SN2021agco}, we constrain the explosion parameters by modeling the multicolor light curves. We provide a discussion in Section \ref{sec: Discussion} and summarize our conclusions in Section \ref{sec: Conclusion}.

\section{Discovery and Observational Properties}\label{sec: Observation_properties}

\subsection{Discovery and Observations}
\begin{figure*}
    \centering
    \includegraphics[width=1\textwidth]{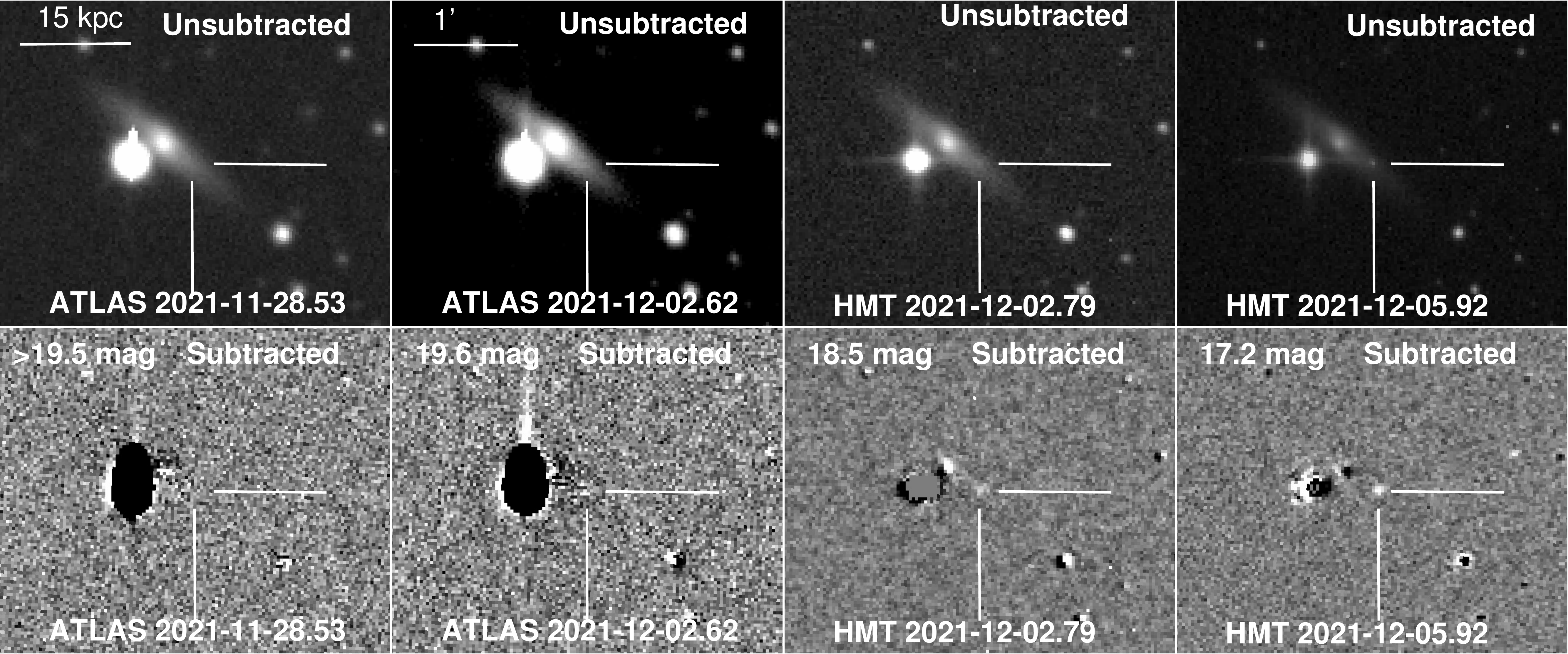}
    \caption{{\it Upper panels:} Early-phase images of the SN 2021agco field observed by ATLAS and the Half Meter Telescope (HMT). {\it Lower panels:} The residual images with host-galaxy template subtracted from the observed ones. In the left two columns are the $o$-band images taken by ATLAS, and the right two are the HMT unfiltered images. The images in the first column represent the nondetections, while the second and third columns represent the first detection. The measured magnitude in each image is labeled on the top left in the lower panels. The nondetection limit of ATLAS on 28.53 Nov. 2021 is $> 19.5$~mag (5$\sigma$). The scale bar represents the scale of the corresponding distance at the host-galaxy redshift.}
    \label{fig: image}
\end{figure*}

SN 2021agco was discovered by Xing Gao on 5 December 2021 at 13:32:13.920 (UTC dates are used throughout this paper; MJD = 59553.56) using a 0.5~m telescope (Half Meter Telescope, dubbed HMT) of Xingming Observatory located at the Nanshan Station of Xinjiang Astronomical Observatory. 
Its coordinates are $\alpha = 07^{\rm hr}28^{\rm m}10.557^{\rm s}$, $\delta = +58^\circ30'12.37''$ (equinox 2000.0), located about $19.6''$ west and $11.4''$ south of the center of the host galaxy UGC~3855. Two survey images taken by HMT at the time of discovery and 3 days before (2 Dec. 2021) are shown in the right two columns of Figure \ref{fig: image}, with the observed and host-subtracted images in the upper and lower panels, respectively. For comparison, the left two columns of Figure \ref{fig: image} display the images taken by ATLAS at $\sim 6$ days (28 Nov. 2021) and 3 days (2 Dec. 2021) before the discovery.We use \texttt{hotpants} to perform the subtraction on the HMT image. From the subtracted images of both ATLAS and HMT, one can see that the earliest detection of SN~2021agco can be traced back to 2 Dec. 2021. We triggered follow-up spectroscopy and photometry of this SN immediately after its discovery.

Photometric observations of SN 2021agco were collected with several telescopes, including the Tsinghua University-NAOC 0.8~m telescope \citep[TNT;][]{Huang_2012, Wang_2008ApJ}, and Ningbo Bureau of Education and Xinjiang Observatory Telescope (NEXT\footnote{\href{ }{http://xjltp.china-vo.org/next.html}}) of Xingming Observatory. The observations of the ATLAS $o$-band (orange) survey had covered the site of the SN, which are also included in our analysis \citep{Tonry_2018}. 
Spectra were obtained with the Beijing Faint Object Spectrograph and Camera (BFOSC) mounted on the Xinglong 2.16~m telescope of NAOC \citep[XLT;][]{XL216m_2016PASP..128k5005F}, the Kast double spectrograph (Kast) on the Shane 3~m telescope at Lick Observatory (Shane), the Yunnan Faint Object Spectrograph and Camera (YFOSC) on the Lijiang 2.4~m telescope (LJT), and the Asiago Faint Object Spectrograph/Camera (AFOSC) on the 1.82~m Copernico Telescope.  Detailed descriptions of observations and data reduction are presented in Appendix \ref{subsec: photo_obs}-\ref{subsec: spec_obs}.

\subsection{Distance and Reddening}
For SN 2021agco, the redshift of its host galaxy UGC 3855 is derived to be $0.010564\pm 0.000027$ from the measurement of the 21~cm line \citep{1998A&AS..130..333T}. Assuming a standard cosmological model and adopting a Hubble constant of 70~$\mathrm{km}\,\mathrm{s}^{-1}\,\mathrm{Mpc}^{-1}$ and $\Omega_m=0.27$, the distance to UGC 3855 can be estimated as 45.6~Mpc and the distance modulus is 33.31$\pm$0.15~mag. Alternatively, the Tully-Fisher relation in three ($JHK$) bands gives a mean distance of $45.0 \pm 9.0$~Mpc and a distance modulus of $33.3\pm0.4$~mag \citep{2007A&A...465...71T}. As the above two methods provide a similar distance modulus, the Tully-Fisher result is adopted throughout this paper.

We estimate the host-galaxy reddening through the Na~I~D $\lambda\lambda$5890, 5896 absorption lines in the observed SN spectra. Considering the quality of the spectra, the $t \approx +5.5$~day Shane/Kast spectrum was used for this purpose, yielding an equivalent width (EW) of $1.13 \pm 0.28$~\AA. Following the relation $E(B-V)=0.16 \times \mathrm{EW({Na~I~D})}$ \citep{EBVvsEW}, this gives ${\rm E(B-V)}_{host}=0.18\pm0.05$~mag. The Galactic reddening toward SN 2021agco is estimated as $E(B-V)_{\rm Galactic} = 0.057$~mag \citep{Galactic_extinction} and the total reddening toward SN~2021agco is thus adopted to be $E(B-V)_{\rm total}$= 0.24$\pm$0.05~mag.

It should be noted that the observed wavelength of the Na~I~D absorption line is not precisely consistent with $(1+z)5893$~\AA; a redshift of $\sim 5$~\AA\ (corresponding to 250 \kms) seems to exist in our spectra, likely caused by the kinematics of disk rotation. UGC 3855 is an edge-on Sab galaxy and the maximum rotation velocity of the gas is derived as $220 \pm 10$ \kms\ through 21~cm measurements \citep{1998A&AS..130..333T}, consistent with the $\sim 5$~\AA\ redshift inferred from our spectra.

\subsection{Light-Curve Properties}\label{subsec: LC_Properties}
\begin{figure*}
    \centering
    \includegraphics[width=1.0\textwidth]{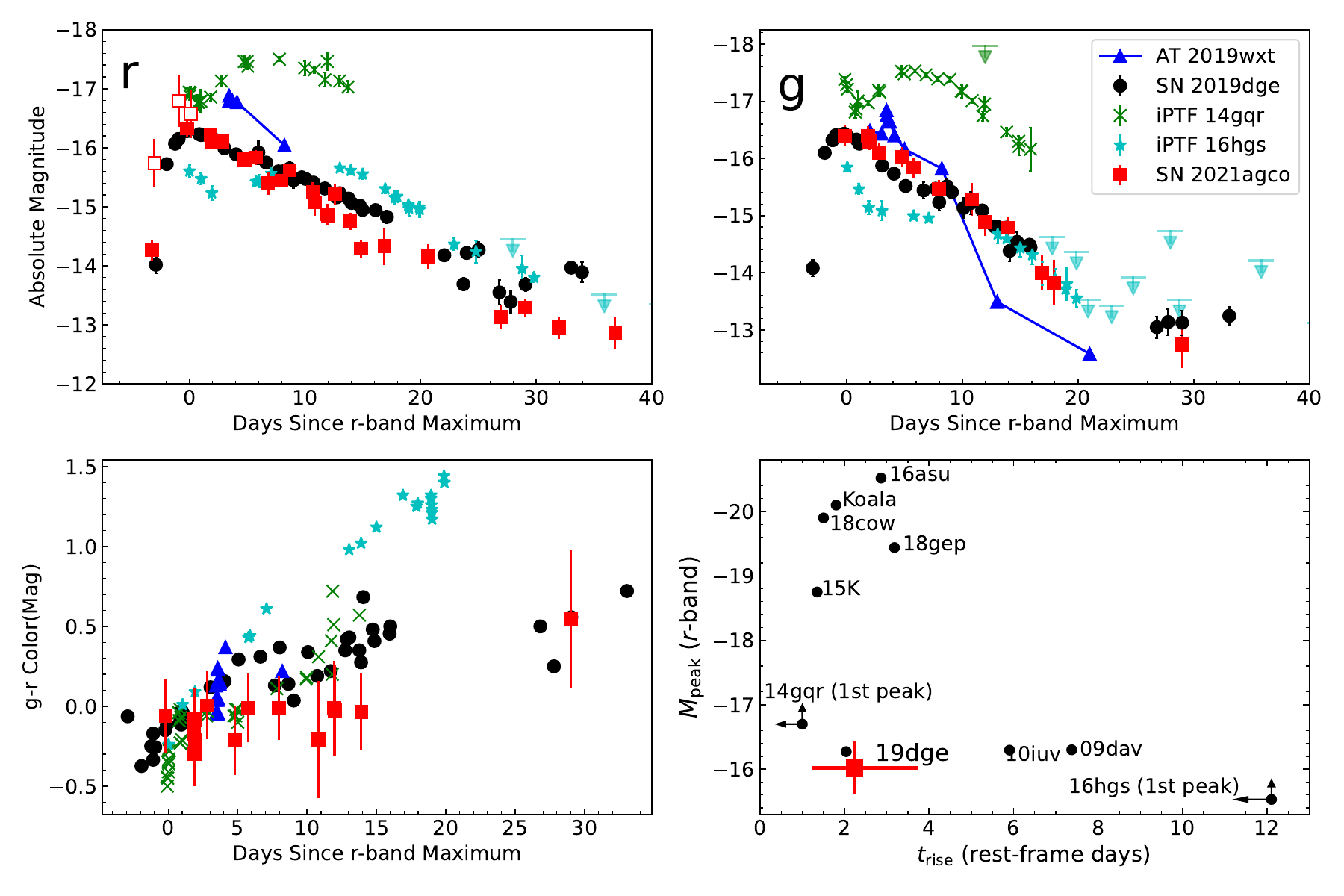}
    \caption{Comparisons of absolute $g$- and $r$-band light curves, $g-r$ color, and rise time of SN 2021agco with other SESNe. {\it Top panels:} The left panel presents the SN 2021agco light curves in $r$ (circles), $o$ (crosses), and clear band (hollow triangles)  with $E(B-V)=0.11$ compared with other SESNe in $r$, while the right panels are in $g$. The epochs correspond to the $r$-band peaks. {\it Bottom-left panel:} SN~2021agco $g-r$ color evolution compared with other SESNe. {\it Bottom-right panel:} The $r$-band peak luminosity vs. the rise time ($t_{\rm rise}$). Comparisons of the decay timescale are shown in Figure \ref{fig: decay_scale}. The objects including fast-evolving Type I transients SN 2002bj \citep{SN2002bj_Poznanski_2010Sci}, SN 2005ek \citep{SN2005ek_Drout_2013ApJ}, SN 2010X \citep{SN2010X_Kasliwal_2010ApJ}, SN 2018kzr \citep{SN2018kzr_McBrien_2019ApJ}, and SN 2019bkc \citep{SN2019bkc_chen_2020ApJ}; Ca-rich transients iPTF16hgs \citep{iPTF_16hgs_De_2018ApJ}, PTF 09dav \citep{PTF09dav_Sullivan_2011ApJ}, and PTF10iuv \citep{PTF10iuv_Kasliwal_2012ApJ}; FBOTs KSN2015K \citep{KSN2015K_Rest_2018NatAs}, iPTF2016asu \citep{iptf16asu_Whitesides_2017}, AT~2018cow \citep{AT2018cow_Prentice_2018, AT2018cow_Perley_2019}, SN 2018gep \citep{SN2018gep_Ho_2019}, and ZTF 18abvkwla \citep{ZTF18abvkwla_Ho_2020}; and ultrastripped SN iPTF14gqr \citep{iPTF14gqr_De_2018} and SN 2019dge \citep{SN2019dge_2020ApJ}.}
    \label{fig: LC_properties}
\end{figure*}

With the above distance and extinction, we get the absolute $g$- and $r$-band light curves of SN 2021agco as shown in Figure \ref{fig: LC_properties}. One can see that this SN has comparable peak magnitudes and shows overall similar light-curve evolution to the first spectroscopically identified helium-rich USSN,  SN 2019dge \citep{SN2019dge_2020ApJ}. In our analysis of the $r$-band evolution, especially at early times, we include the ATLAS $o$-band and HMT clear-band data because of their similarities to SDSS-$r$ in wavelength coverage. See Appendix \ref{sec: LC_correction} and Appendix \ref{subsec: LC_Interpolation} for detailed photometric correction and light-curve fitting. The combined data indicate that this SN rose very rapidly immediately after the explosion. In comparison with the luminosity distribution of normal SNe~Ib/Ic, the absolute $r$-band peak magnitude ($\sim -16.06\pm0.42$~mag) of SN 2021agco is at the lower end \citep{Drout_2011ApJ, Taddia_2018A&A}.

In Figure \ref{fig: LC_properties}, we also plot the $g-r$ color of SN 2021agco together with that of SN 2019dge and SN~2019wxt to examine the temperature evolution. After peak brightness, SN 2021agco is as blue as the other two USSNe, while it is found to maintain a constant color of about 0.1--0.2~mag for at least 2 weeks. Subsequently, the color evolution is similar to that of SN 2019dge.

Following the definition of rise time by \cite{SN2019dge_2020ApJ}, namely the time that it takes from half of the peak to the peak brightness, we calculate $t_{\rm rise}$ to be $\sim 2.4^{+1.5}_{-1.0}$~days for SN 2021agco and the decay timescale $t_{\rm decay}$ to be $\sim 7.6$~days. In Figure \ref{fig: LC_properties}d, we compare $t_{\rm rise}$ and $r$-band peak absolute magnitude of SN 2021agco with those of other fast-evolving transients identified over the past few years, including FBOTs like AT~2018cow and AT~2018gep and some calcium-rich transients like iPTF10iuv and iPTF09dav. The comparison data are mainly taken from \cite{SN2019dge_2020ApJ}. The rise time of SN~2021agco is much shorter than that of typical SNe~Ib/Ic \citep{Taddia_2015A&A} and the calcium-rich transients as well, while its absolute peak magnitude is much fainter than that of typical FBOTs. 

\subsection{Bolometric Light Curves}\label{sec: BLC}
Using the $BVgrizo$-band photometry, we construct the spectral energy distribution (SED) and bolometric light curve of SN 2021agco, along with the effective temperature and photospheric radius. The comparisons of the above parameters among the USSN candidates are shown in Figure \ref{fig: Bolometric_LC}. For fitting details see Appendix \ref{subsec: BLC_construction}. 
 
As can be seen from Figure \ref{fig: Bolometric_LC}, SN 2021agco shows a similar pseudobolometric light curve as SN~2019dge and SN~2019wxt within 2 weeks after peak brightness. The temperature might decrease a few days after the peak, and it then becomes constant at $\sim 8000$~K. The photosphere tends to expand continuously to $\sim 7 \times 10^{3}$~\rsun\ until $\sim 3$~days after the peak and it then begins to slowly recede. The photospheric evolution of SN 2021agco is more similar to that of SN 2019dge, while other USSN candidates expand to larger radii ($>20\times10^3$~\rsun) until $\sim 15$~days, followed by a more rapid receding.

\begin{figure}
    \centering
    \includegraphics[width=0.45\textwidth]{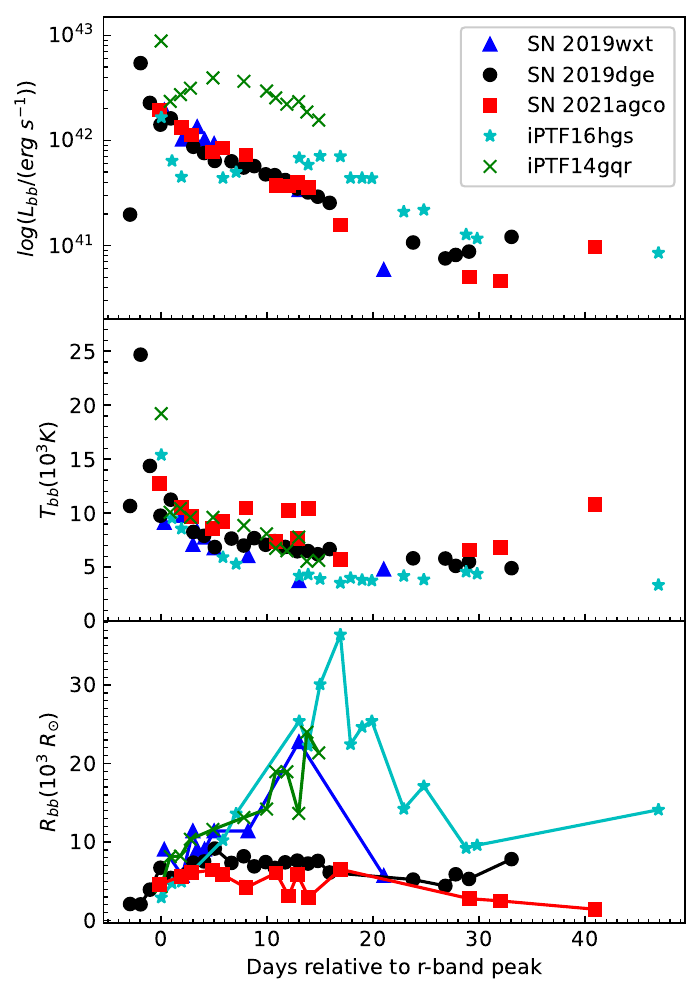}
    \caption{The evolution of the pseudobolometric light curve, photospheric temperature, and blackbody radius of SN 2021agco compared with those of other known USSN candidates, including SN~2019dge, SN~2019wxt, iPTF14gqr, and the Ca-rich gap transient iPTF16hgs. The phases are relative to the $r$-band maximum as a reference for each transient. The bolometric light curves of SN~2019dge and SN~2019wxt are taken from \cite{SN2019dge_2020ApJ} and \cite{AT2019wxt_2022arxiv}, respectively. For iPTF14gqr and iPTF16hgs, the adopted construction methods are the same as for SN 2021agco. }
    \label{fig: Bolometric_LC}
\end{figure}

\subsection{Optical Spectral Properties}\label{sec: Optical Spectral Properties}
\begin{figure*}
    \centering
    \includegraphics[width=1\textwidth]{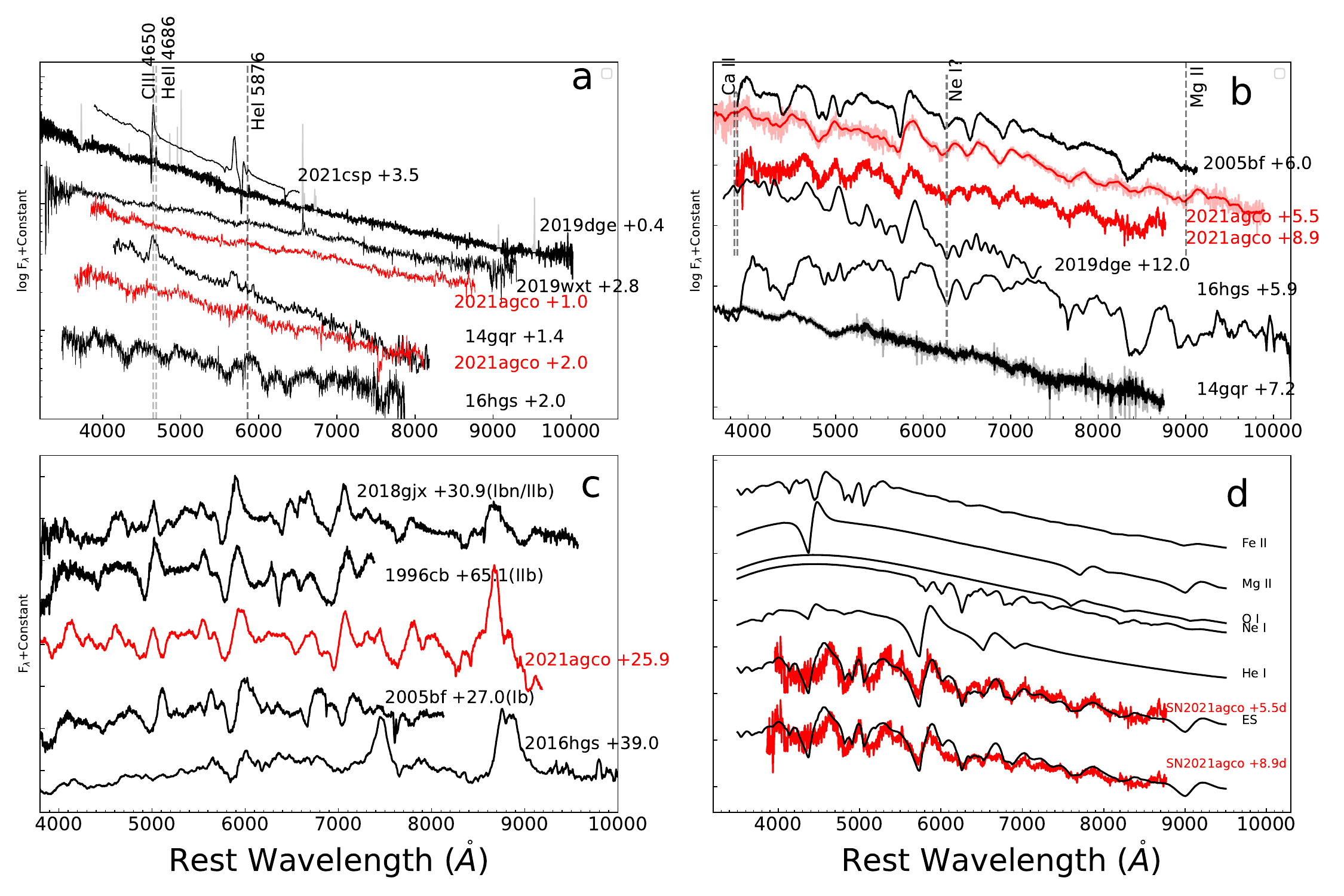}
    \caption{Panel (a): $t \approx +1.0$ and +2.0~day spectra of SN 2021agco (red lines) compared with peak spectra of other SESNe (black lines). Strong emission lines from the host galaxy are marked with gray lines. Panel (b): Spectra at +5.5 and +8.9 days post peak compared with SN 2005bf at +6 days and SN 2019dge at +11.8 days . Panel (c): SN 2021agco spectrum at +25.9 days compared with other helium-rich SNe. The flux density is plotted on a linear scale and adjusted for better display. The spectra of SN 2021agco and iPTF16hgs have been rebinned to improved the spectral quality. Panel (d): The  spectra taken at +5.5 and +8.9 days (red lines) compared with the synthetic spectra (black lines at the bottom). The best-fit SYN++ model spectra together with absorption lines due to different ions are shown with black lines.}
    \label{fig: compare_spec}
\end{figure*}

The detailed spectroscopic comparisons between SN 2021agco and other USSNe are shown in Figure \ref{fig: compare_spec}, where the first three panels provide comparisons of the spectra at peak brightness, 1--2 weeks after peak, and 1--2 months after peak, respectively (see Figure \ref{fig: all_spectra} for the  spectral series of SN~2021agco, covering the phases from +1.0 to +25.9~days after the $r$-band peak).

The earlier spectra ($t \approx +1.0$ and +2.0 days) are characterized by blue continua with shallow P~Cygni profiles of \ion{He}{1} $\lambda5876$, which is similar to that of USSN candidates. The weak emission line at $\sim 4650$~\AA\ is visible, which could be identified as \ion{C}{3} $\lambda4650$. With the photosphere receding into the deeper ejecta shell, the P~Cygni profile of \ion{He}{1} lines becomes progressively stronger. In the $t \approx 5.5$ and 8.9~day spectra, absorption lines of \ion{Fe}{2}, \ion{Mg}{2}, and \ion{Ne}{1} become prominent, and are further confirmed by SYNAPPS \citep{synapps} fitting (see details in Appendix \ref{subsec: SYN++} and Figure \ref{fig: compare_spec}d). The photospheric spectra are identical to those of the Type Ib SN 2005bf \citep{SN2005bf_anupama_sn_2005, SN2005bf_tominaga_unique_2005, SN2005bf_folatelli_sn_2006, SN2005bf_parrent_direct_2007}, and share several similar absorption features with SN 2019dge. By $t \approx 1$ month, emission features of the Ca~II near-infrared (NIR) triplet seem to appear in the spectrum, although the spectral quality is low. Nevertheless, the presence of prominent \ion{He}{1} lines and the absence of \ion{H}{1} lines can safely classify SN 2021agco as a helium-rich SN. 

\section{Modeling of SN 2021agco}\label{sec: modeling of SN2021agco}

In this section, we model the multicolor light curves to investigate the explosion parameters of SN 2021agco. The light-curve data include the ATLAS-$o$ and $BVgriz$ data from TNT and NEXT. We apply the Monte Carlo Markov Chain (MCMC) sampling method to estimate the best parameters through \texttt{emcee} \citep{emcee_Foreman_Mackey_2013}. According to the results in Section \ref{subsec: LC_Properties}, the explosion time is fixed as $\mathrm{MJD}_{\rm exp}=59550.36 \pm 0.06$.

\subsection{Nickel-Powered Model}

The thermalization of radioactive \Ni decay has been widely and successfully used to explain the light curves of normal SNe~Ib/Ic. The bell-shaped light curve is consistent with a low ejecta mass and a small radius of the progenitor \citep{Taddia_2018A&A}. Given that SN~2021agco has a single light-curve peak, we first attempt to reproduce its multicolor light curves using the \citet{Arnett_1982ApJ} model. The detailed definitions of photospheric radius and temperature are described in Appendix \ref{app: Ni_model}. The opacity $\kappa_{\rm opt}$ is adopted as 0.07~cm$^2$~g$^{-1}$, which is taken as the mean value of SNe~Ib/Ic \citep{Taddia_2018A&A}.

The best-fitting parameters are given as $E_{k}= 7.5_{-0.04}^{+0.11}\times10^{48}$~erg for the kinetic energy, $M_{\rm ejecta}=0.04$~\msun\ for the ejecta mass, and $M_{\rm Ni} \approx 0.03$~\msun\ for the synthesized nickel mass. In the Arnett model, the rapid rise of the early-time light curve indicates a short diffusion time, corresponding to a small amount of ejecta mass. Meanwhile, sufficient nickel mass is needed to power the light-curve peak, which results in a very high fraction of \Ni mass in the ejecta. However, a larger \Ni fraction in ejecta would lead to stronger iron-line blanketing, which is not seen in the spectra. The unusually low ejecta mass, high fraction of \Ni mass, and poor fitting motivate us to explore a more reasonable model.

\subsection{Shock Cooling}

The above analysis shows that the single component of the radioactivity-powered model cannot reproduce the rapid rise of the early-time light curve of SN 2021agco. Therefore, the fitting needs to consider a new energy source and radiative-transfer model. The early-time spectra of SN 2021agco present featureless blue continua, with distinct absorption lines emerging in subsequent spectra. This indicates that the electrons are rapidly captured by ions and the hot plasma suffers fast cooling. The USSNe SN 2019dge \citep{SN2019dge_2020ApJ} also exhibits such a signature, which was interpreted as shock-cooling emission of the photosphere \citep{SCE_Piro_2021}. Thus, we also include a shock-cooling component in the model fitting in addition to the radioactive-decay component.

In the fitting, both components are simultaneously fit to match the observed multicolor light curves. The best-fit results are shown in Table \ref{tab:fitresultsebv}. Combining the results, the suggested properties of the progenitor envelope are as follows: an envelope radius of $R_{\rm env}\sim78.4_{-19.9}^{+25.6}$ \rsun, a mass of $M_{\rm env}=0.10_{-0.01}^{+0.02}$ \msun, and an injection energy of $E_{\rm ext}=8.93_{-1.61}^{+2.59}\times10^{49}$~erg. The SN ejecta have a mass of $M_{\rm ejecta}=0.26_{-0.02}^{+0.04}$ \msun, a \Ni mass of $M_{\rm Ni}=2.2_{-0.3}^{+0.2} \times 10^{-2}$ \msun, and kinetic energy of $E_{\rm k}=9.57_{-1.62}^{+3.01}\times10^{49}$~erg.

\begin{figure*}
    \centering
    \includegraphics[width=1\textwidth]{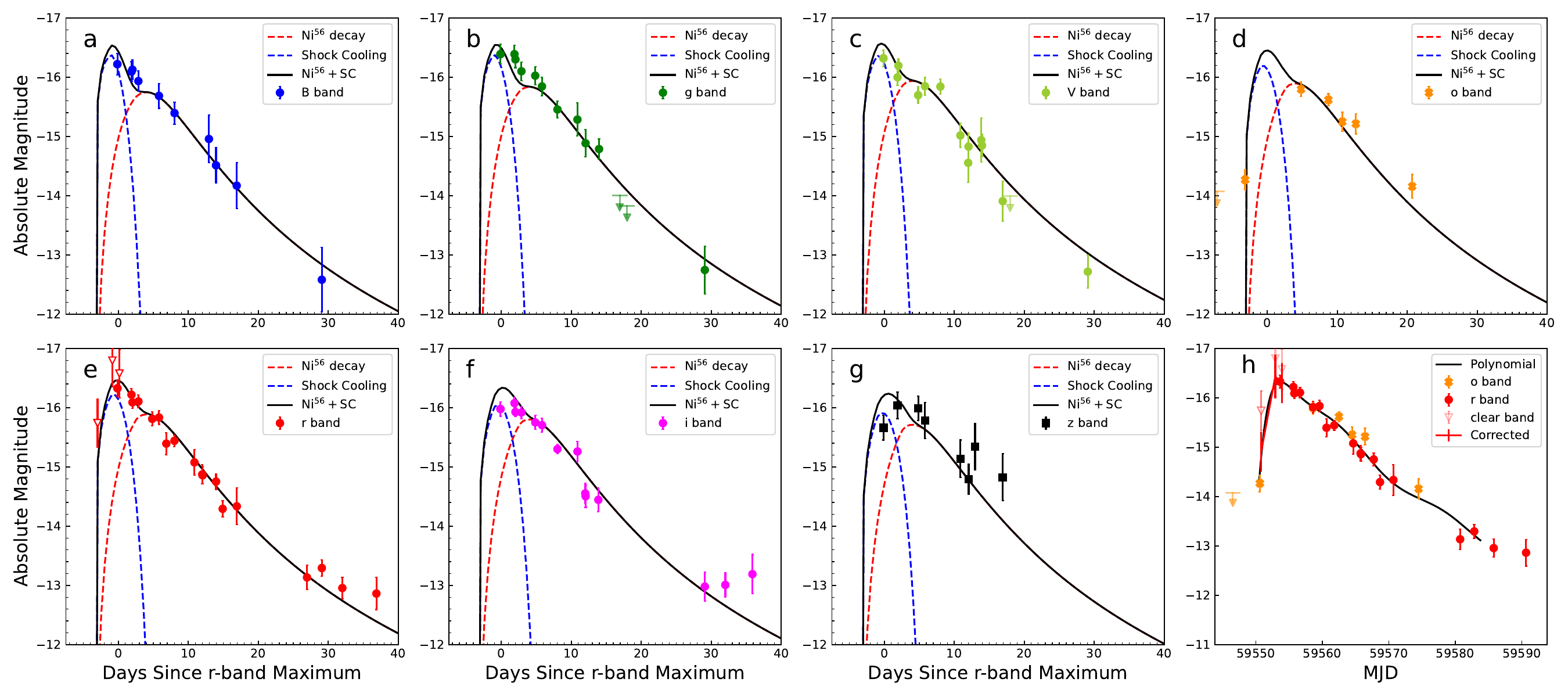}
    \caption{Comparison of the observed $BgVoriz$-band light curves of SN 2021agco with the models considering energy contribution from shock cooling and Ni$^{56}$ decay. A total reddening of E$(B - V)$ = 0.24 mag is applied to the extinction corrections of all observed light curves.  Panels a-g show the observed luminosity compared with the best-fit model light curves in each band. The data are taken by NEXT (squares), TNT (circles), ATLAS (orange crosses), and HMT (hollow red triangles). The blue dashed line and red dashed line represent the shock-cooling component and the radioactive decay component, respectively. While the black lines represent the total luminosity. Notice that the $z$-band light curve is not included in the light-curve fitting, while the modeled light curves are still displayed in panel $g$. Panel $h$ displays the HMT clear-band, TNT \& NEXT $r$-band, and ATLAS $o$-band photometry compared with the light curve modeled by a polynomial function (gray line).}
    \label{fig: LC}
\end{figure*}

\subsection{Circumstellar Medium}
The circumstellar interaction (CSI) model can also reproduce the light curves of diverse peak luminosities and fast-evolving transients \citep{Pellegrino_2022ApJ}. Very weak emission lines might be visible in the early-time spectra of SN 2021agco but disappear in subsequent spectra. Some  FBOTs, such as AT~2018cow, lack prominent narrow emission lines in their early-time spectra, while the light curves still had been successfully interpreted with the CSI model \citep{Xiang_2018cow_2021ApJ}. So, the absence of persistent emission lines may not be direct evidence against the CSI model; it requires further investigation on this subclass of USSNe but is beyond the scope of this article.

Highly-ionized narrow emission lines (i.e., the blending features at $\sim$4600 \AA) have been detected in the early spectra of He-poor USSNe iPTF 14gqr \citep{iPTF14gqr_De_2018} and the He-rich USSNe including SN~2019dge \citep{SN2019dge_2020ApJ} and SN~2019wxt \citep{AT2019wxt_2022arxiv}. Such a narrow emission line is also detected  at $\sim$4650~\AA\ in the t$\sim$+1.0 day spectrum of SN 2021agco. This feature is measured to have a width of $\sim20$~\AA\ and a flux of 1.4$\times$10$^{-13}$~erg~cm$^{-2}$ s$^{-1}$, corresponding to a luminosity of 4.1$\times$10$^{38}$ erg~s$^{-1}$ and a velocity of $\sim$1300 km s$^{-1}$. Assuming that this spectral line is caused by highly ionized carbon, i.e., C~III~$\lambda4650$, with the electron density being adopted as $n_e=3n_{C^{+++}}$ and the radiative rates (A-value) being as $A=2.2\times 10^{-24}~erg~{\rm cm}^{3}~{\rm s}^{-1}$ \citep{2015MNRAS.450.1151A}, the radius of the CSM is estimated as $\gtrsim$280\rsun, and the mass loss rate of the CSM is $\gtrsim0.006$~\msun~$\text{yr}^{-1}$. 
Assuming the $\sim$4600 \AA\ emission feature as He II would lead to an unusually large blueshift velocity for the CSM (i.e., $\sim$2500 km s$^{-1}$), which was rarely seen in the ionized-flash spectra of SNe.

\section{Discussion}\label{sec: Discussion}
\subsection{Host Galaxy}\label{sec: Host Galaxy}
The host of SN 2021agco is classified as an Sab galaxy\footnote{Refer to NASA/IPAC Extragalactic Database (NED) \href{ }{http://ned.ipac.caltech.edu}}. The SN site is at a projected distance of about 22$^{\prime\prime}$ from the galactic center, corresponding to a projected distance of $\sim 4.8$~kpc. To further investigate the properties of its host galaxy, we took a spectrum near the center of UGC 3855 with the Lick 3~m Shane telescope (+Kast), as shown in the bottom panel of Figure \ref{fig: all_spectra}. The absorption features of Mg~I $\lambda$5175, Na~I~D, and TiO are visible, while a narrow \Ha\ line is barely visible. We apply a spectral fitting code using the stellar population synthesis technique \citep[\texttt{FIREFLY};][]{FireFly_1st_2017MNRAS.472.4297W, FireFly_MaStar_2020MNRAS.496.2962M} to examine the host-galaxy properties of SN 2021agco. The best fit gives a metallicity of $\sim 1.3$~\zsun, a stellar population age of 10.0~Gyr, a host-galaxy mass of $M_{\rm galaxy} = 2.6\times10^{9}$~\msun, and a light-weighted star-formation rate (SFR) of $\sim 0.2$~\msun~yr$^{-1}$. Note, however, that the spectrum of the SN 2021agco host galaxy was taken at its central region, which tends to hold older stellar populations. While the SN site lies within the disk of the  galaxy, where the properties of the stellar population might differ significantly from that at the galactic center. Therefore, an accurate population environment cannot be established for the progenitor of SN 2021agco. Additional observations and larger samples are required to ascertain whether all USSNe originate from young stellar populations.

In comparison with normal SNe~Ib, the host galaxies of USSNe are found to have relatively lower metallicity. For example, the host galaxy of SN~2019dge has a metallicity of $\sim 0.4$~\zsun\ \citep{SN2019dge_2020ApJ}. For both SN 2021agco and iPTF14gqr, the host galaxies are found to have a metallicity of 1.2--1.3~\zsun\ \citep{iPTF14gqr_De_2018}, while the host galaxies of normal SNe~Ib/Ic have metallicity in the range 2.5--2.8~\zsun\ \citep{Schulze_SNeHost_2021ApJS}.

\subsection{Comparison with Other USSNe Candidates}
\input{Exp_param_USSNe.tex}

The explosion and host-galaxy parameters of SN 2021agco and a few USSN candidates are shown in Table \ref{tab: Exp_param_USSNe}. Owing to extremely rapid post-explosion evolution, early-time observations are usually absent for them, leading to larger uncertainties in their light-curve modeling. Therefore, we can only make some qualitative comparisons between them at some times. 

{\bf SN 2019dge:} SN 2021agco shares many similarities with SN 2019dge. In comparison, SN 2021agco has broader and deeper spectral absorption features, which might indicate a higher fraction of helium in its ejecta. Moreover, the fitting shows that SN 2021agco has a smaller envelope but more-massive ejecta, indicating a denser progenitor envelope. Combining the above characteristics, SN 2021agco might originate from a relatively higher-mass helium star in a short-period binary system, as suggested by \cite{tauris_USSN_2015}.

{\bf iPTF14gqr:} Compared with the two confirmed Type Ib USSNe SN 2019dge and SN 2021agco, another USSN possible candidate is iPTF14gqr, the second peak of whose light curve is broader and more luminous, and occurs at a later phase of $t \approx 1$ week after the explosion. A prominent emission feature near 4600~\AA\ is visible in the $t \approx 0.5$--1.5 day spectra of iPTF14gqr (see Figure \ref{fig: compare_spec}a), which can be a blend of \ion{C}{3} $\lambda$4650 and \ion{He}{2} $\lambda$4686. Note, however, that this feature in iPTF14gqr appears much stronger and broader than that seen in SN 2021agco, SN 2019dge, and SN~2019wxt, which may be due to the helium-rich circumstellar medium (CSM) shell around iPTF14gqr being accelerated by a shock. Moreover, we notice that iPTF14gqr shows similar photospheric radius evolution as SN~2019wxt, both expanding and reaching the largest size at $t \approx 2$--3 weeks after the explosion.

{\bf SN~2019wxt:} The light curve of SN~2019wxt is found to decline faster than that of other USSN candidates, especially in the $g$ band (see the top-right panel in Figure \ref{fig: LC_properties}). As explosion parameters suggested by \cite{AT2019wxt_2022arxiv}, the ejecta and \Ni mass of SN~2019wxt are higher than in SN 2021agco, yet the light curves decline faster. The higher ejecta and \Ni mass may be caused by the poor limits on the explosion time. The constructed bolometric luminosity at $t \approx 20$ days is brighter than observed, indicating an overestimation of the \Ni or ejecta mass. The photospheric radius of SN~2019wxt shows similar evolution to the Type Ic USSN candidate iPTF14gqr, reaching the maximum value at $t \approx 2$ weeks after explosion, while it differs from SN 2021agco and SN 2019dge which seem to remain at a constant photospheric radius. 

\subsection{Progenitor of SN 2021agco}
By fitting the multicolor light curves of SN 2021agco, one can see that the ejecta and envelope masses of the progenitor are lower than those of typical SESNe \citep{Taddia_2018A&A}. For single-star models, mass loss via stellar winds cannot be responsible for such a violent mass stripping. So, interaction with the companion star in a binary system is required for mass loss during stellar evolution. As the nickel mass derived for SN 2021agco ($\sim 0.02$~\msun) is larger than that predicted by electron-captured SNe \citep[$M_{\rm Ni}\sim10^{-3}$~\msun,][]{Moriya_ECSN_2014A&A}, and the spectral line velocity (7000~\kms) is also faster than that of fallback SNe \citep[i.e., $\sim3000$~\kms;][]{Valenti_2008ha_2009Natur, Moriya_fallbackSNe_2010ApJ}, we propose that SN 2021agco should be an iron core-collapse SN (Fe-CCSN)

The absence of H features in SN 2021agco indicates that most of the hydrogen layer was stripped away by the companion \citep{Laplace_2020A&A}. To further peel the outer layer of a primary stars, a short orbital period is required for the binary system \citep{Yoon_2010ApJ, Long_2022}. If the binary system goes through a common-envelope evolution, it will finally evolve into a He-star--NS binary with a close orbit \citep{Dewi_2003MNRAS, CCE_Ivanova_2013A&ARv}. For the helium star with $M_{\rm He} \lesssim 4$~\msun, after core helium depletion the star expands and triggers Roche-lobe overflow (RLO), so-called Case~BB RLO \citep{1986A&A...165...95H, 1986A&A...167...61H}. Then, the compact NS in the tight binary system would further strip the outer layer of the helium donor star, finally leading to an Fe-CCSN \citep{tauris_USSN_2013, tauris_USSN_2015}. \cite{Ca-Rich_Zenati_2023ApJ} and \cite{SN2019ehk_Jacobson_2020ApJ} suggested a model of hybrid HeCO+CO white dwarf system, where the Helium shell detonated partially on the surface of the low mass primary ($\lesssim~0.65$\msun), which well explained the observed features of some Ca-rich transients. But the main observed features of SN~2021agco do not match those of the Ca-rich transients. 

The spectra of SN~2021agco show remarkable differences from those of Ca-rich SNe. The emission feature of Ca II/O II at $\sim$ 7300 \AA\ is invisible in SN~2021agco spectra, while it is very prominent in those Ca-rich SNe. SN 2021agco was  apparently bluer than the Ca-rich SNe after t$\sim$1 week from the maximum light. For example, at t$\sim$2 weeks from the peak, the $g-r$ color is measured as 0.4-0.5 mag for SN 2021agco while it is 1.1-1.2 mag for iPTF16hgs (a ca-rich SN). Finally, SN 2021agco shows a more rapid luminosity evolution, with the rise time being only about 2-3 days, while the typical rise time for the ca-rich SNe is about 2 weeks. Assuming a thermonulcear origin, a rise time of 2-3 days corresponds to a peak luminosity that is much lower than -14.0 mag in B \citep[see Fig. 7 of ][]{Ca-Rich_Zenati_2023ApJ}, which is inconsistent with that seen in SN 2021agco. Thus, we exclude the possibility of SN~2021agco as a member of Ca-rich transients.
\section{Conclusion}\label{sec: Conclusion}

We present optical observations and analysis of an extremely rare stellar explosion, SN 2021agco. This object represents the third ultrastripped SN of the Type Ib subclass. The light curve shows very fast evolution, reaching the peak of $M_{\rm peak} \approx -16.06\pm0.42$~mag in the $r$ band within 3 days after the explosion. The early-time spectrum shows a featureless blue continuum with a narrow, weak emission line that can be attributed to \ion{C}{3} $\lambda$4650, while the later spectra exhibit prominent helium lines with P~Cygni profiles. By comparison, we find that both the light curves and spectra of SN 2021agco show close resemblances to those of SN 2019dge except that the former have broader line profiles. 

We fit the light curve of SN 2021agco by using a joint model of shock-cooling emission and radioactive decay of \Ni. With the MCMC fitting method, we derived the SN ejecta mass $M_{\rm ejecta}=0.26_{-0.02}^{+0.04}$~\msun, \Ni mass $M_{\rm Ni}=2.2_{-0.3}^{+0.2} \times 10^{-2}$~\msun, and the kinetic energy $E_{\rm k}=9.57_{-1.62}^{+3.01}\times10^{49}$~erg. The progenitor is estimated to have an envelope with a radius $R_{\rm env}\sim78.4_{-19.9}^{+25.6}$~\rsun, a mass $M_{\rm env}=0.10_{-0.01}^{+0.02}$~\msun and an injection energy $E_{\rm ext}=8.93_{-1.61}^{+2.59}\times10^{49}$~erg. The less-massive ejecta and envelope indicate that the progenitor suffered violent mass loss and the majority of its outer layer was stripped before the explosion. 

We further examined the properties of the host galaxy of SN 2021agco by analyzing its spectrum, finding that it is a relatively old (10.6~Gyr) Sab galaxy with a metallicity of 1.3~\zsun\ and a star-formation rate of 0.2~\msun~yr$^{-1}$. Because the spectra were obtained at the center of the galaxy, the environment of the old stellar population is not expected for USSNe. Additional observations and larger samples are required to ascertain whether all USSNe originate from young stellar populations.

\begin{acknowledgements}
We acknowledge the support of the staff of the XLT, LJT, TNT, HMT, and Lick 3~m Shane telescope for assistance with the observations. The operation of XLT is supported by the Open Project Program of the Key Laboratory of Optical Astronomy, National Astronomical Observatories, Chinese Academy of Sciences. Funding for the LJT has been provided by the Chinese Academy of Sciences and the People's Government of Yunnan Province. The LJT is jointly operated and administrated by Yunnan Observatories and Center for Astronomical Mega-Science, CAS.
		
This work is supported by the National Natural Science Foundation of China (NSFC grants 12288102, 12033003, and 11633002), the Scholar Program of Beijing Academy of Science and Technology (DZ:BS202002), and the Tencent Xplorer Prize. A.V.F.'s group at U.C. Berkeley received financial support from the Christopher R. Redlich Fund, Frank and Kathleen Wood (T.G.B. is a Wood Specialist in Astronomy), and many individual donors. J.J.Z. and Y.Z.C. are supported by the International Centre of Supernovae, Yunnan Key Laboratory (No. 202302AN360001). Y.Z.C. is supported by the National Natural Science Foundation of China (NSFC, Grant No. 12303054).

\end{acknowledgements}

\appendix

\section{Optical Photometric and Spectroscopic Observations}\label{app: obs_log}

\subsection{Photometric Observations}\label{subsec: photo_obs}

Follow-up photometry of SN 2021agco was collected with several telescopes, including the Tsinghua University-NAOC 0.8~m telescope \citep[TNT;][]{Huang_2012, Wang_2008ApJ} at Xinglong Observatory of NAOC, and the Ningbo Bureau of Education and Xinjiang Observatory Telescope (NEXT\footnote{\href{ }{http://xjltp.china-vo.org/next.html}}) of Xingming Observatory. 

The TNT observations were obtained in standard Johnson-Cousins $BV$ bands and SDSS $gri$ bands, while the NEXT observations were obtained in $BV$ and SDSS $griz$ bands. The clear-band photometry derived from the HMT survey images and the $o$-band (orange) data collected by the ATLAS telescope system on Haleakala in Hawaii, USA \citep{Tonry_2018, 2020PASP..132h5002S} are also included in our analysis. 
\input{Photomety}

The photometric images were preprocessed using standard IRAF\footnote{IRAF is distributed by the National Optical Astronomy Observatories, which are operated by the Association of Universities for Research in Astronomy, Inc., under a cooperative agreement with the U.S. National Science Foundation (NSF).} routines. A Python-based automatic photometric data-processing pipeline \citep[\texttt{AutoPhOt};][]{Autophot_pipeline} is adopted in our data reduction. As SN 2021agco is not very bright and is near the host-galaxy center, template subtraction is needed for accurate photometry. The TNT and NEXT template images were taken on 18.5 April 2022 (MJD = 59687.5) and 24.7 Jan. 2022 (MJD = 59634.7), respectively, when the target became faint enough. For clear-band photometry from the HMT survey, we used the image taken on 27 Nov. 2021 as templates. We use the catalogs of Pan-STARRS\footnote{\href{ }{https://panstarrs.ifa.hawaii.edu/pswww/}} and APASS to calibrate the $griz$-band and $BV$-band photometry, respectively. Details of the photometric measurements and upper limits obtained from the ATLAS, HMT, TNT, and NEXT are shown in Table \ref{tab:photometry}. 

\subsection{Optical Spectroscopy}\label{subsec: spec_obs}
We obtained four low-resolution optical spectra of SN 2021agco, spanning the phases from $t \approx +1.0$ to $t \approx +25.9$ days relative to $r$-band maximum light. The spectra were collected by several instruments, including the Beijing Faint Object Spectrograph and Camera (BFOSC) mounted on the Xinglong 2.16~m telescope of NAOC \citep[XLT;][]{XL216m_2016PASP..128k5005F}, the Kast double spectrograph (Kast) on the Shane 3~m telescope at Lick Observatory (Shane), and the Yunnan Faint Object Spectrograph and Camera (YFOSC) on the Lijiang 2.4~m telescope (LJT). The above spectra were reduced using standard IRAF pipeline and Python/IDL codes\footnote{\href{ }{https://github.com/ishivvers/TheKastShiv}}. Note that a $t \approx +2.0$ days spectrum posted on the TNS \citep{2021TNSCR4118....1T}, obtained with the 1.82~m Copernico Telescope and the Asiago Faint Object Spectrograph/Camera (AFOSC), is also included in our analysis. The spectra taken by XLT, Kast, and LJT were obtained using the parallactic angle \citep{Filippenko_1982} and with airmass $\lesssim$1.2. The log of the five spectroscopic observations is given in Table \ref{tab: spectra log}. All spectra are shown in the upper panel of Figure \ref{fig: all_spectra}. 
\input{SpecTable}

\section{Light-Curve Corrections and Properties}
\subsection{Photometric Correction of HMT}\label{sec: LC_correction}
Since the ATLAS $o$, HMT clear, and TNT $r$ bands have similar central wavelengths, we combine the above photometric data to produce the $r$-band light curve. First, we need to correct the HMT clear-band photometric system compared to the $r$ band. But owing to the lack of color and temperature evolution at early times, while SN 2021agco shares many similar observational characteristics with SN 2019dge, we assume the temperature SN 2021agco evolves similarly to that of SN 2019dge. Next, we fit the SN 2019dge temperature using a toy model, $T=a/(t-t_0)+b$, where $t_0$, $a$, and $b$ are free parameters. The fit result is shown in Figure \ref{fig:Scorrection}(a). With the reproduced temperature evolution, we interpolate the observation time corresponding to the HMT. After interpolation, we take the temperature of the corresponding phases of HMT ($\sim 67,000$, 22,000, and 13,000~K). Since the early-time spectrum of SN 2021agco is almost a featureless blue continuum, the pre-peak spectra are also assumed to be single blackbodies. Then we synthesize the magnitude by convolving an instrumental response function $[S(\lambda)]$ with the blackbody spectra inferred above. The synthesizing function is given as

\begin{equation}
  \mathrm{mag = -2.5log} \int~ N_{\lambda}S(\lambda) + \mathrm{ZP}\, ,
\end{equation}
\noindent
where the $N_{\lambda}$ is the photon spectrum and ZP is the zero point of $S(\lambda)$. The response functions $S(\lambda)$ include the atmosphere extinction of observatories, the filter transmissions, and the quantum efficiency (QE) of the detectors. Such information for HMT, TNT, and ATLAS is shown in the top three panels of Figure \ref{fig:Scorrection}(c1-c3), while the last panel displays the combination of these effects. Note that $S(\lambda)$ may also be affected by mirror reflectance and other unknown optical losses. However, owing to a lack of this kind of information, we did not consider these effects in $S(\lambda)$. We convolve the spectra and response curve and calculate the zero point using \texttt{pyphot}\footnote{\href{ }{https://mfouesneau.github.io/pyphot/index.html}}. Using the HMT, TNT, and ATLAS photometry systems, after convoluting the blackbody spectra at different temperatures, we can calculate the discrepancies in the magnitudes measured by these instruments. Finally, we corrected the HMT photometric system into the TNT $r$ band. The corrected HMT results are shown in Figure \ref{fig:Scorrection}(b).

\subsection{Interpolation}\label{subsec: LC_Interpolation}
\begin{figure*}
    \centering
    \includegraphics[width=1.0\textwidth]{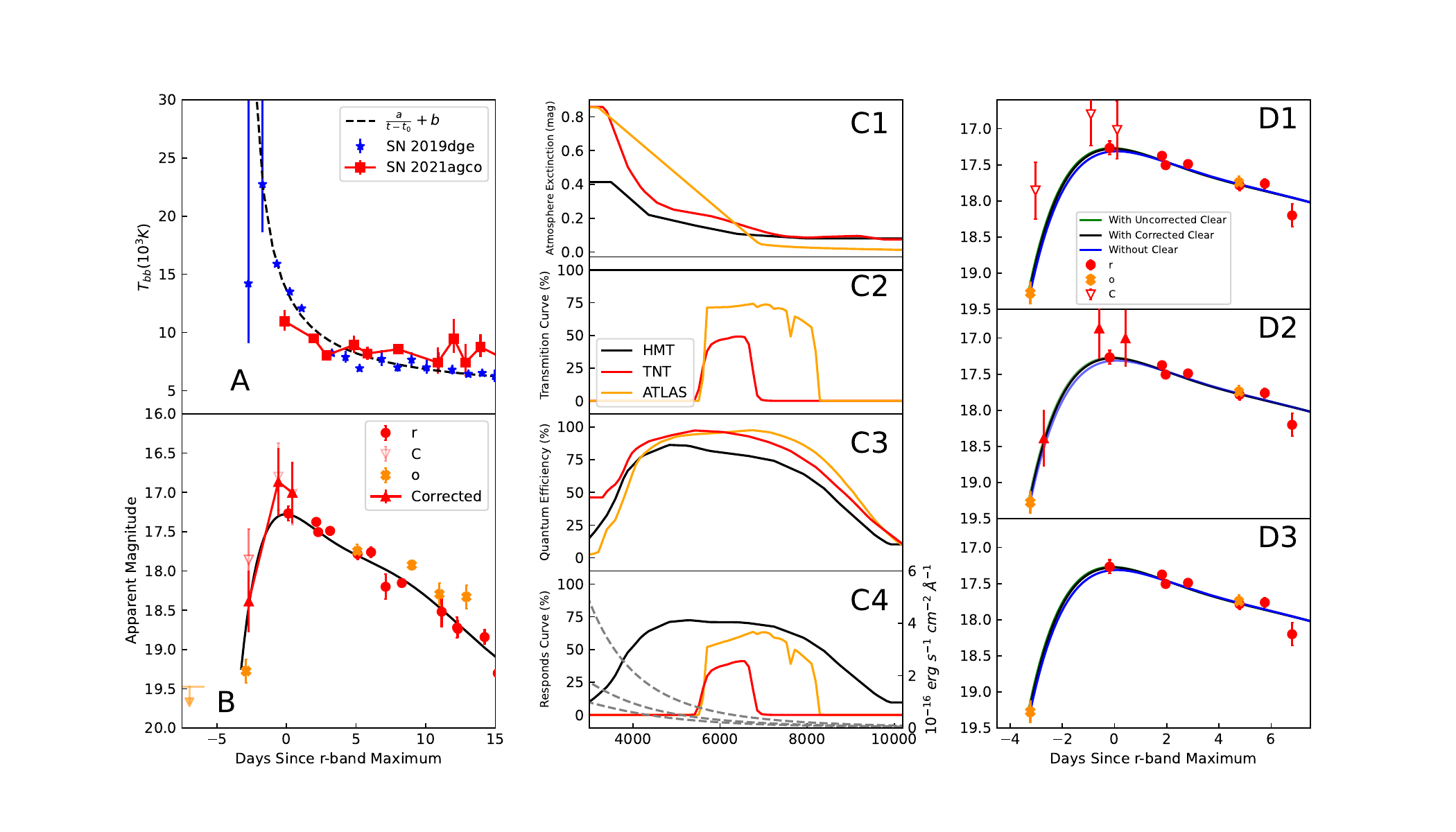}
    \caption{The correction for the HMT photometric data. {\it Panel~a:} The blue stars are the temperature evolution of SN 2019dge taken from \cite{SN2019dge_2020ApJ}. The black dotted line is a toy model to interpolate the SN 2019dge temperature evolution, which is expressed as $T=a/(t-t_0)+b$. The red square is the temperature evolution of SN 2021agco, which is derived from Section \ref*{sec: BLC}. {\it Panel~b:} The red points and orange crosses are SDSS $r$-band and ATLAS $o$-band data, respectively. The translucent red triangles are the uncorrected HMT clear-band photometric data, while the connected red triangles are the corrected ones. The black line is a polynomial function to fit the joint data of the ATLAS $o$, corrected HMT clear, and $r$ bands. {\it Panel~c1-c4:} Panel~c1-c3 three panels display the atmosphere extinctions, transmission curves of filters, and CCD quantum efficiency of HMT, TNT, and ATLAS, respectively. Panel c4 shows the response curves of the combinations of the above three effects. {\it Panel~d1-d3:}: The symbolic representation of the observed data is consistent with that of panel B. In addition to combining SDSS $r$-band and ATLAS data, we combined uncorrected HMT data, corrected HMT data, and without HMT data for polynomial fitting. Panels d1-d3 show the fitting results of the above three cases, respectively.
    }
    \label{fig:Scorrection}
\end{figure*}

To better quantify the evolution timescale and brightness, we need to fit the observed data with an analytical function. We fit the TNT, ATLAS, and corrected HMT photometric data using a polynomial function, which is shown in Figure \ref{fig:Scorrection} (d2). Meanwhile, when fitting the polynomial functions, TNT and ATLAS data are still included. We also consider both cases with uncorrected HMT data and without HMT data, which are shown in Figure \ref{fig:Scorrection} (d1 and d3). We can see that, because of the large uncertainties of the HMT photometry, the fitting results are slightly affected by the HMT. 

\begin{figure}
    \centering
    \includegraphics[width=0.5\textwidth]{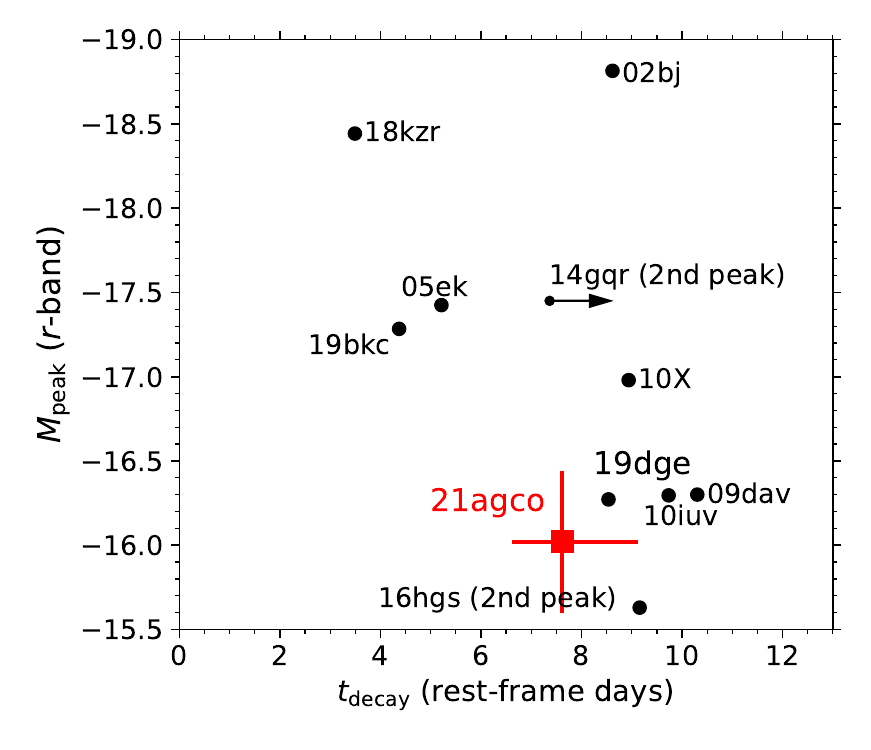}
    \caption{The comparison of the light curve decay timescale of the fast-evolving transients. The included transients are the same as in the last panel in Figure \ref{fig: LC_properties}.}
    \label{fig: decay_scale}
\end{figure}

\begin{figure*}
    \centering
    \includegraphics[width=1\textwidth]{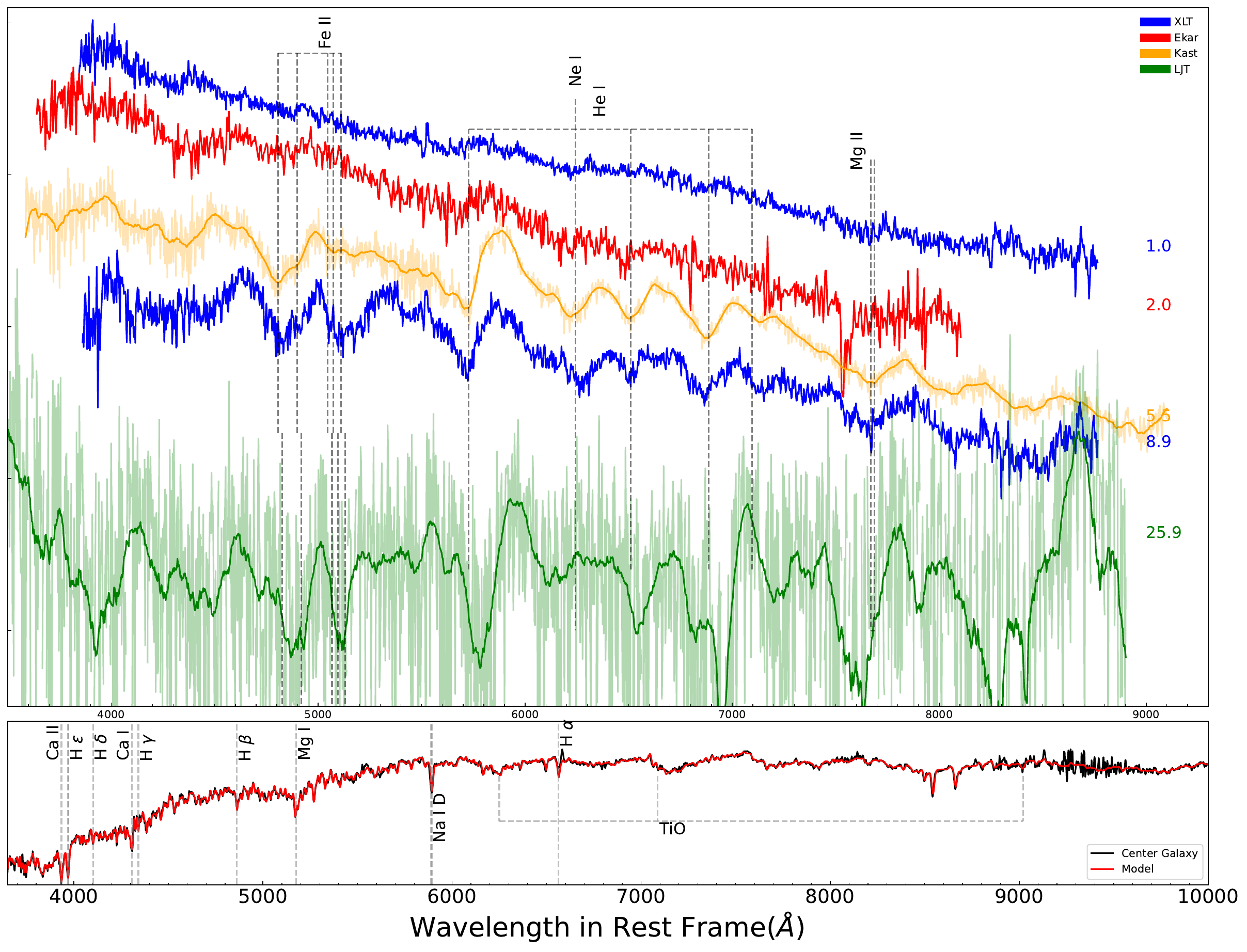}
    \caption{{\it Upper panel:} Optical spectra of SN2021agco. The spectral wavelength is corrected for the host-galaxy redshift, $z=0.010564$. The flux density is plotted on a logarithmic scale and shifted for better display. Different colors indicate the spectra taken with different telescopes: XLT (blue), Copernico Telescope (red), Shane (orange), and LJT (green). The two spectra obtained with the Lick 3~m Shane telescope and the Lijiang 2.4~m telescope are smoothed using the Savitzky Golay method with a window size of 61 and polynomial order of 2, with the original spectra shown in translucent colors. The number marked on the left of each spectrum indicates the phase relative to the $r$-band maximum. {\it Lower panel:} The spectrum of the center of the host galaxy UGC 3358 of SN2021agco (black lines) compares with a FIREFLY modeled spectrum (red lines). The absorption features are marked by gray dashed lines.}
    \label{fig: all_spectra}
\end{figure*}

Finally, we adopt the fit result which contains the corrected HMT data to calculate the SN 2021agco peak magnitude and its time in the $r$ band, which is $M_{r,{\rm peak}} = -16.06\pm0.42$~mag at $\mathrm{MJD} =59553.56$ days. Using the fitted polynomial results, we could also calculate the rise and decay timescales, which are defined as a rise from 0.75~mag below the peak to the peak, and a decline from the peak by 0.75~mag (0.75~mag below peak means half of the peak flux). The SN 2021agco rise timescale compared with other fast-involving transients is shown in the last panel in Figure \ref{fig: LC_properties}, and the decay timescale is shown in Figure \ref{fig: decay_scale}.

When determining the explosion date, we join the first observation of ATLAS and HMT (MJD $< 59551$), and adopt a function of luminosity evolution as $L_{\lambda} \propto (t - t_{\rm exp})^n$, where $n$ was fixed at the typical value of 2. The derived explosion date is $\mathrm{MJD}_{\rm exp}=59550.36 \pm 0.06$.

\section{Spectral Properties}
\subsection{Peak Spectra}
At around peak brightness, SN 2021agco is characterized by a blue continuum with shallow spectral features, including absorption lines of \ion{He}{1} $\lambda\lambda$5876, 6678, 7065, \ion{Fe}{2} absorption in the range 4200--4500~\AA, and the \ion{Ca}{2} NIR triplet. In the $t \approx +1.0$ day spectrum, a weak emission feature is barely visible at $\sim 4650$~\AA; it could be due to \ion{C}{3}~$\lambda4650$ and \ion{He}{2}~$\lambda4686$, However, we notice that its full width at half-maximum intensity (FWHM) is consistent with the spectral resolution ($\sim 21$~\AA, corresponding to $\sim 1400$~\kms), suggesting that it is a single emission line without contamination. As shown in the upper panel in Figure \ref{fig: compare_spec}, such a feature is also visible in the early-time spectra of several comparison USSNe. It was identified as a blend of \ion{C}{3}~$\lambda4650$ and \ion{He}{2}~$\lambda4686$ in iPTF14gqr \citep{iPTF14gqr_De_2018}, \ion{He}{2} in SN 2019dge, and \ion{C}{3}~$\lambda4650$ in SN~2019wxt \citep{AT2019wxt_2022arxiv} and SNe~Icn such as SN 2021csp \citep{SN2021csp_Morgan_2021arXiv, SN2021csp_Perley_2022ApJ}. In SN 2021agco, we identify this emission feature as \ion{C}{3}~$\lambda4650$, as its central wavelength is at $\lambda4650$. (However, if one assumes it is actually \ion{He}{2}~$\lambda4686$, it would have a blueshifted velocity of $\sim 2500$~km~s$^{-1}$.)

The weak emission line at $\sim 4650$~\AA\ is likely produced by a carbon-enriched CSM shell. In comparison, a weak \ion{C}{3}~$\lambda$4650 emission feature seems to be also visible in the $t \approx +2.75$~day spectrum of SN 2019wxt \citep{AT2019wxt_2022arxiv}, which is measured to have a FWHM velocity of $\sim 4300$~\kms. The larger velocity of SN~2019wxt may be due to line blending with \ion{He}{2}~$\lambda$4686, or perhaps its CSM has been shocked at this phase. In SN 2019dge, however, prominent narrow \ion{He}{2}~$\lambda4686$ emission can be identified in the early spectra covering phases from $t \approx -1.1$~day to +0.4~day. 

As the ionization energy of \ion{C}{3}~$\lambda4650$ and \ion{He}{2}~$\lambda4686$ are similar (24.4~eV vs. 24.6~eV), the nondetection of narrow \ion{He}{2}~$\lambda4686$ in SN 2021agco seems unreasonable. This absence of \ion{He}{2}~$\lambda4686$ may be related to the late observed phase. As shown in Figure 14 of \cite{tauris_USSN_2015}, when the SN exploded, the helium was in the outer layer while the carbon was in the inner layers. The outer layers expand faster and cool down more quickly. The electrons recombined with \ion{He}{2} and the corresponding emission faded away more quickly, while the carbon in the inner layer remained hot, producing weak emission lines. Alternatively, such a difference may be related to the composition of the surrounding CSM. The carbon-enriched CSM of SN 2021agco may be similar to that of the recently discovered Type Icn SNe \citep{SN2019hs_Gal_Yam_2022Natur, SN2021csp_Morgan_2021arXiv, SN2021csp_Perley_2022ApJ, SN2022ann_2022Davis}.

\subsection{Photospheric Spectra}

At this phase, broader \ion{He}{1} features are visible in $t \approx 5.5$~day and $t \approx 11.9$~day spectra of SN 2021agco, including the \ion{He}{1}~$\lambda\lambda$5876, 6678, 7065 lines. \ion{Fe}{2} lines near 4900~\AA\ and the \ion{Ca}{2} NIR triplet become prominent in the spectra at this phase. Note that both spectra of SN 2021agco are found to have a deep absorption near 6200--6270~\AA, which is also visible in the SN 2019dge and iPTF16hgs spectra at similar phases. Such an absorption feature exists in the $t \approx -1$~day spectrum of the peculiar Type Ib SN 2005bf. Although high-velocity \Ha\ and/or \ion{Si}{2} have been proposed in earlier literature \citep{SN2005bf_anupama_sn_2005, SN2005bf_tominaga_unique_2005, SN2005bf_folatelli_sn_2006, SN2005bf_parrent_direct_2007}, a unanimous identification of this feature is absent. Our SYNAPPS fit suggests the absorption feature near 6200--6270~\AA\ can be well explained by \ion{Ne}{1}~$\lambda6402$ (see Section \ref{sec: spec_V}). The overall spectral features of SN 2021agco are similar to those of SN 2005bf but at a slightly earlier phase \citep{SN2005bf_anupama_sn_2005, SN2005bf_parrent_direct_2007, SN2005bf_folatelli_sn_2006}.

With the help of the SN spectral identification tool GELATO \citep{Gelato_2008A&A}, the $t \approx +25.9$~day spectrum of SN 2021agco, with a prominent P~Cygni profile of \ion{He}{1}~$\lambda5876$, is found to match well with later spectra of two SNe~IIb: SN 1996cb at $t \approx 65$~day \citep{SN1996cb_1999AJ} and SN 2005bf at $t \approx 45$~day. This suggests that SN 2021agco has a faster spectral evolution than other He-rich SNe. Moreover, the fact that the absorption at $\sim 6300$~\AA\ still exists in SN 1996cb and SN 2008bo but disappears in SN 2021agco indicates that such an absorption may have different origins among them. SN~Ibn and SN~IIb transitional supernova SN 2018gjx  \citep{Prentice_2018gjx_2020MNRAS} shares some similar absorption lines with SN 2021agco at the same epoch. We also compare the spectrum of SN 2021agco with that of Ca-rich gap transient iPTF16hgs and found that its Ca~II NIR triplet tends to grow stronger quickly at a later phase.

\subsection{Velocity of Spectral Lines}\label{sec: spec_V}
\begin{figure*}
    \centering
    \includegraphics[width=1\textwidth]{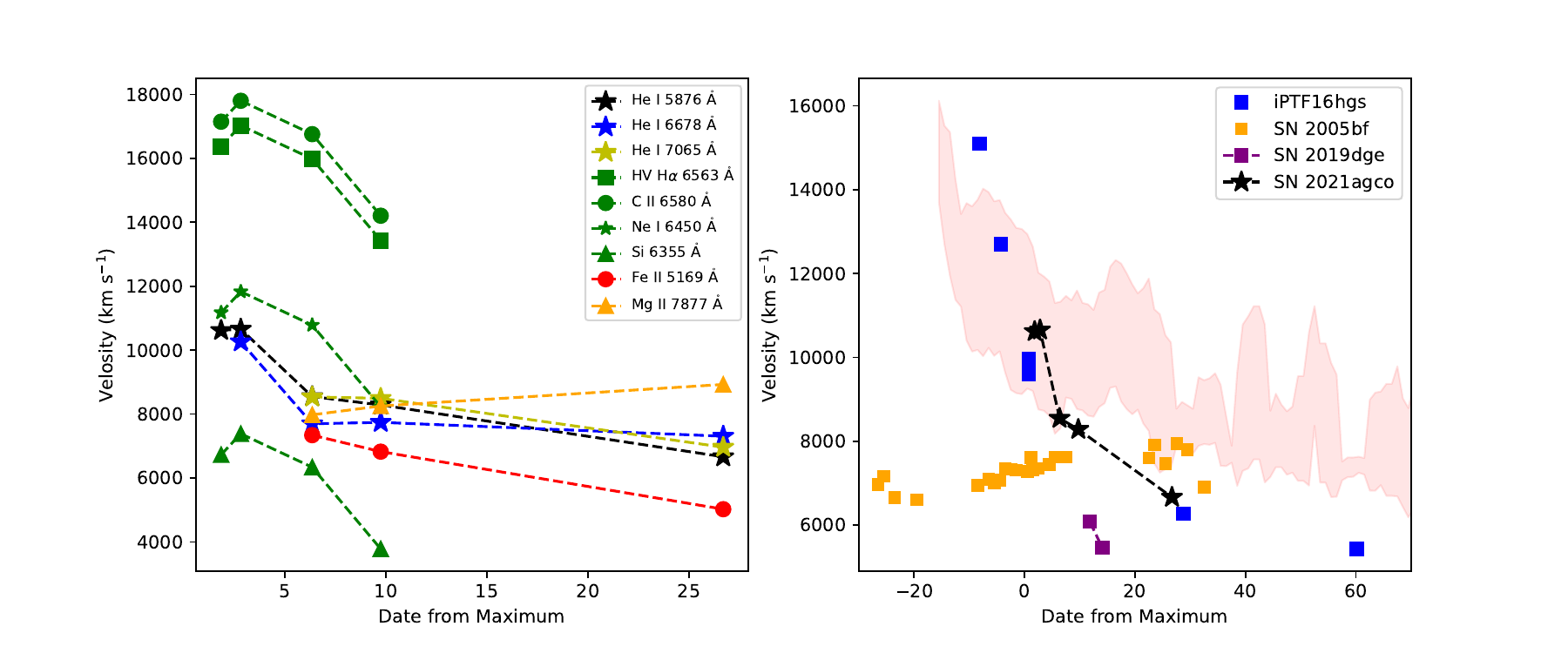}
    \caption{{\it Left:} Line velocity evolution of \ion{He}{1}~$\lambda\lambda 5876$, 6678, 7065, \ion{Mg}{2}~$\lambda7877$, and \ion{Fe}{2}~$\lambda5169$ in the spectra of SN 2021agco, labeled in different colors (see legend). As inferred from the absorption at $\sim 6300$~\AA, the green dashed lines connecting different symbols are the assumed velocities of \Ha, \ion{C}{2}~$\lambda6580$, \ion{Ne}{1}~$\lambda6450$, and \ion{Si}{2}~$\lambda6355$. {\it Right:} \ion{He}{1}~$\lambda5876$ spectral line velocity of SN 2021agco compared with other Type Ib SNe. The red region is the rolling weighted average values for SNe~Ib, which are taken from  \cite{Ibc_spec_velocity_Liu_2016}.}
    \label{fig: LineVelocity}
\end{figure*}

To examine the evolution of spectral lines, we measure the velocity inferred from absorption minima of \ion{He}{1}~$\lambda\lambda$5876, 6678, 7065, \ion{Mg}{2}~$\lambda7877$, and \ion{Fe}{2}~$\lambda5169$, as shown in the left panel in Figure \ref{fig: LineVelocity}. To identify the absorption at $\sim 6300$~\AA, the central wavelength is measured and assumed to be \Ha, \ion{C}{2}~$\lambda6580$, \ion{Ne}{1}~$\lambda6450$, and \ion{Si}{2}~$\lambda6355$, whose inferred velocities are displayed in green and labeled with different shapes. After the peak, the \ion{He}{1} velocity is found to decline rapidly from $\sim 10,000$ to 8000~\kms\ within about 10 days. Assuming the absorption as \Ha, however, the inferred expansion velocity (i.e., $\sim 13,000$--16,000~\kms) is much higher than velocities deduced from other elements \citep{Ibc_spec_velocity_Liu_2016}. The velocity of the assumed \ion{Si}{2}~$\lambda$6355 is much lower than the photospheric velocity inferred from \ion{Fe}{2}~$\lambda5169$ \citep{Photosphere_Dessart_2015MNRAS.453.2189D, Photosphere_Fe5169_Modjaz_2016}, while \ion{C}{2} is much higher. The velocity of \ion{Ne}{1} $\lambda$6402 is more consistent with that of other elements.

The left panel in Figure \ref{fig: LineVelocity} shows the comparison of velocity evolution of \ion{He}{1}~$\lambda5876$ for SN 2021agco, iPTF16hgs, SN 2019dge, and SN 2005bf. The average velocity evolution obtained for SNe~Ib and SNe~IIb are overplotted for comparison \citep{Ibc_spec_velocity_Liu_2016}. The velocity of SN 2021agco and iPTF16hgs is much higher than that of SN 2019dge. One can see that both SN 2021agco and iPTF16hgs show much faster velocity evolution than normal SNe~Ib/IIb, suggesting that the photosphere recedes into the inner region at a faster pace.

\subsection{SNAPPS Fitting}\label{subsec: SYN++}

To better identify the elements existing in the ejecta of SN 2021agco, we apply the spectral synthesis code SYNAPPS  \citep{synapps} to reproduce the $t \approx 5.5$ day (11 Dec. 2021) and $t \approx 10.5$~day (14 Dec. 2021) spectra of SN 2021agco. In the fitting, the optical-depth profile was assumed as an exponential format, while the continuum temperature and photospheric velocity were adopted as 9000~K and 7000~\kms, respectively, for both spectra. Comparison between the observed and modeled spectra favors the presence of \ion{He}{1}, \ion{Ne}{1}, \ion{Mg}{2}, \ion{Ca}{2}, and \ion{Fe}{2} ions, as seen in Figure \ref{fig: compare_spec}d. In particular, the noticeable absorption feature near 6210--6270~\AA\ can be reasonably identified as \ion{Ne}{1}$~\lambda6402$. Including \ion{Ne}{1} absorption features at 6929 and 7032~\AA\ can also help explain the absorption intensity at $\sim 6900$~\AA. So, in our analysis, \ion{Ne}{1}~$\lambda$6450 is adopted for the 6210--6270~\AA\ absorption seen in SN 2021agco.

\section{Light-Curve Modeling}

\subsection{Bolometric Light Curves}\label{subsec: BLC_construction}
We construct the bolometric light curve for SN 2021agco. We assume the SED is a blackbody. We use the MCMC sampling method to estimate the photospheric radius and temperature. The model priors for the photospheric radius and temperature are assumed uniformly and logarithmically distributed within [$10,10^{6}$]~\rsun\ and [$10^{3},10^{6}$]~K, respectively. The fitting results, including bolometric luminosity, photospheric temperatures, and radius with $E(B-V)_{host}=0.06$ are shown in Table \ref{tab:BB_evolution}. 

As can be seen from Figure \ref{fig: Bolometric_LC}, SN 2021agco shows a similar pseudo bolometric light curve as SN 2019dge and SN~2019wxt within 2 weeks after the peak. After that, SN~2019wxt and SN 2021agco seem to decline faster than SN 2019dge. Notice that SN~2019wxt is also found to have an overall relatively lower temperature in comparison with the latter two helium-rich USSNe. For both SN 2021agco and SN 2019dge, the temperature decreased gradually from about 10,000~K a few days after peak brightness to $\sim 6000$~K a few weeks thereafter. The corresponding temperature inferred for SN~2019wxt is lower by about 2000~K at comparable phases. Moreover, the photosphere of these two helium-rich USSNe is found to initially expand within $\sim 5$~days and then maintain a constant radius of $5\times10^{3}$~\rsun. In contrast, the photospheric radius of iPTF14gqr, SN~2019wxt, and iPTF16hgs expands to a larger radius (i.e., $\gtrsim 2\times10^{4}$~\rsun) within $\sim 20$~days and then contracts with time, which is more similar to the behavior of normal SNe~Ib/Ic \citep{Taddia_2018A&A}. 

\subsection{Nickel Radioactive Decay Model}\label{app: Ni_model}

When modeling the multiband light curve, the first step is to define the photospheric radius and temperature. We assume the photosphere expands at a constant velocity at early times. With the ejecta expanding and cooling to a critical temperature, the photospheric temperature would evolve as a constant value ($T_{f}$), and the radius would recede into the inner radius. The relation of luminosity, photospheric temperature, and radius could be calculated from a Stefan–Boltzmann law, given by

\begin{equation}
    \label{equ: Tf}
    T_{\rm phot}\left(t \right) =\left\{ \begin{array}{c}
	\left[ \frac{L\left(t \right)}{4\pi \sigma \left( v_{\rm phot}t \right) ^2} \right] ^{\frac{1}{4}}, \left[ \frac{L\left(t \right)}{4\pi \sigma \left( v_{\rm phot}t \right) ^2} \right] ^{\frac{1}{4}}>T_f\\
	T_f, \left[ \frac{L\left(t \right)}{4\pi \sigma \left( v_{\rm phot}t \right) ^2} \right] ^{\frac{1}{4}}\le T_f\\
\end{array} \right. 
\end{equation}

\begin{equation}
    \label{equ: Rphot}
R_{\rm phot}\left(t \right) =\left\{         \begin{array}{c}
	v_{\rm phot}t, \left[ \frac{L\left(t \right)}{4\pi \sigma \left( v_{\rm phot}t \right) ^2} \right] ^{\frac{1}{4}}>T_f\\
	\left[ \frac{L\left(t \right)}{4\pi \sigma T_f^4} \right] ^{\frac{1}{2}}, \left[ \frac{L\left(t \right)}{4\pi \sigma \left( v_{\rm phot}t \right) ^2} \right] ^{\frac{1}{4}}\le T_f\\
\end{array} \right. 
\end{equation}
\noindent
where the $T_{\rm phot}$, $R_{\rm phot}$, and $v_{\rm phot}$ are the photospheric temperature, radius, and velocity (respectively), $\sigma$ is the Stefan–Boltzmann constant, and $L(t)$ is the total luminosity. The $T_f$ is fixed at 10000~K, which is inferred from Section \ref{sec: BLC}. By the MCMC sampling method, one can estimate the model parameters $M_{\rm Ni}$, $M_{\rm ejecta}$, and $E_k$.

\subsection{Shock Cooling and Radioactive Decay}

\begin{figure*}
    \centering
    \includegraphics[width=1\textwidth]{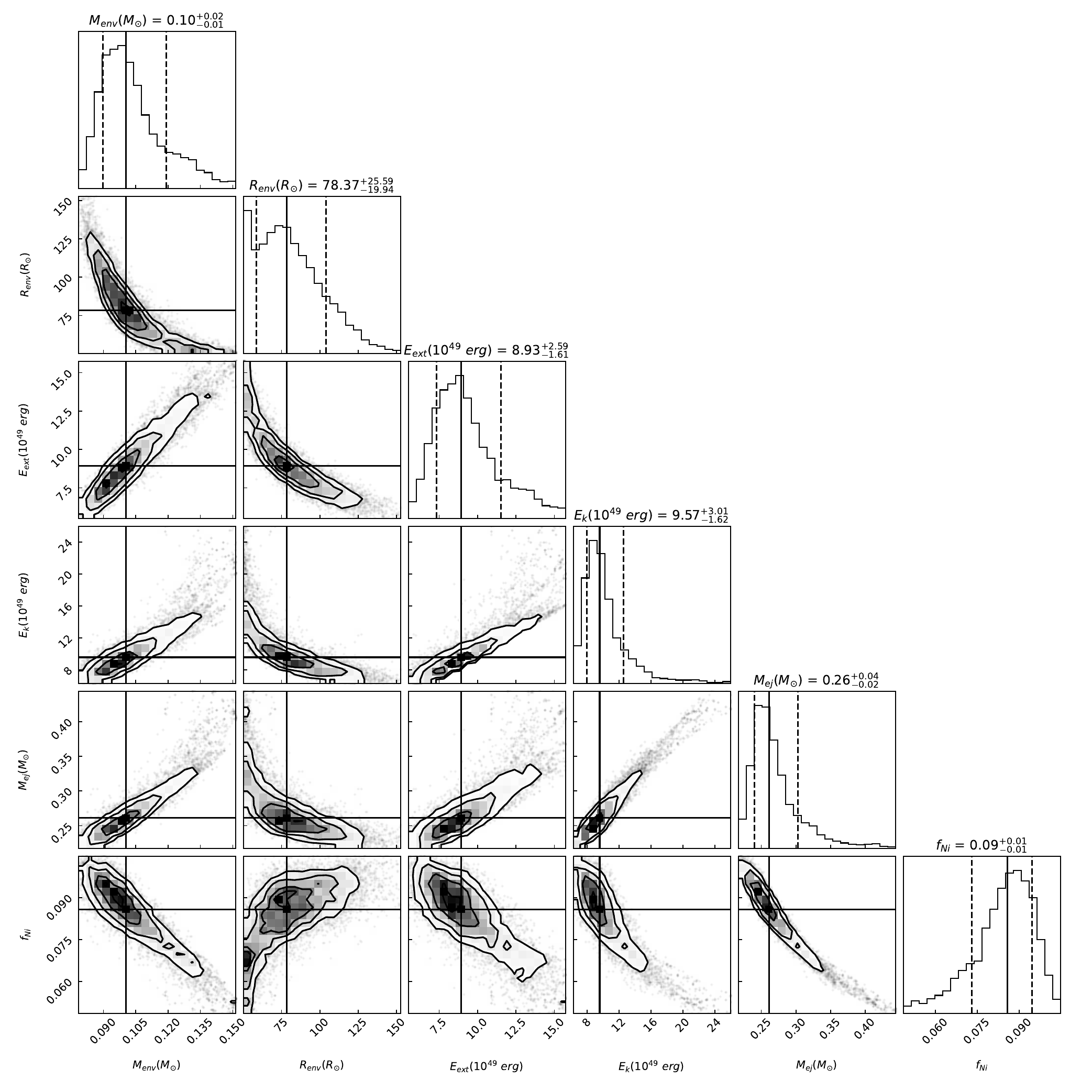}
    \caption{The corner plots shows posterior distributions for the joint model of the shock cooling and \Ni decay.}
    \label{fig: Moedel_coner}
\end{figure*}

The single-component model of \Ni decay cannot well explain the SN 2021agco light curves, which motivates us to explore a new model with an additional power source. A two-component model for shock cooling and \Ni decay is adopted. In this case, the total luminosity can be expressed as 
\begin{equation}
\label{equ: TotalLuminous}
    L_{\rm total}=L_{\rm SCE}+L_{\rm Ni}\, ,
\end{equation}
where $L_{\rm SCE}$ represents the energy contributed by shock-cooling emission  \citep{SCE_Piro_2021} and $L_{\rm Ni}$ represents the contribution by \Ni decay. By using the MCMC sampling method, the posterior distribution of the parameters is displayed in Figure \ref{fig: Moedel_coner}. The fitting results with are presented in Table \ref{tab:fitresultsebv}.
\input{bolometric}

\input{EBV}
\bibliographystyle{aasjournal}
\bibliography{bibtex}

\end{document}

%% file: author_name.tex
\correspondingauthor{Xiaofeng Wang}
\email{wang\_xf@mail.tsinghua.edu.cn}

\author{Shengyu Yan}
\affiliation{Department of Physics and Tsinghua Center for Astrophysics (THCA), Tsinghua University, Haidian District, Beijing 100084, China}

\author{Xiaofeng Wang}
\affiliation{Department of Physics and Tsinghua Center for Astrophysics (THCA), Tsinghua University, Haidian District, Beijing 100084, China}
\affiliation{Beijing Planetarium, Beijing Academy of Sciences and Technology, Beijing 100044, China}

\author{Xing Gao}
\affiliation{Xinjiang Astronomical Observatory, Urumqi 830011, China}

\author{Jujia Zhang}

\affiliation{Yunnan Observatories, Chinese Academy of Sciences, Kunming 260216, China}

\affiliation{Key Laboratory for the Structure and Evolution of Celestial Objects, Chinese Academy of Sciences, Kunming 650216, China} 

\affiliation{International Centre of Supernovae, Yunnan Key Laboratory, Kunming 650216, China}

\author{Alexei V. Filippenko}
\affiliation{Department of Astronomy, University of California, Berkeley, CA 94720-3411, USA}

\author{Thomas G. Brink}
\affiliation{Department of Astronomy, University of California, Berkeley, CA 94720-3411, USA}

\author{Jun Mo}
\affiliation{Department of Physics and Tsinghua Center for Astrophysics (THCA), Tsinghua University, Haidian District, Beijing 100084, China}

\author{Weili Lin }
\affiliation{Department of Physics and Tsinghua Center for Astrophysics (THCA), Tsinghua University, Haidian District, Beijing 100084, China}

\author{Danfeng Xiang}
\affiliation{Department of Physics and Tsinghua Center for Astrophysics (THCA), Tsinghua University, Haidian District, Beijing 100084, China}

\author{Xiaoran Ma}
\affiliation{Department of Physics and Tsinghua Center for Astrophysics (THCA), Tsinghua University, Haidian District, Beijing 100084, China}

\author{Fangzhou Guo}
\affiliation{Department of Physics and Tsinghua Center for Astrophysics (THCA), Tsinghua University, Haidian District, Beijing 100084, China}

\author{Lina Tomasella}
\affiliation{INAF -- Osservatorio Astronomico di Padova, Vicolo dell'Osservatorio 5, 35122 Padova, Italy}

\author{Stefano Benetti}
\affiliation{INAF -- Osservatorio Astronomico di Padova, Vicolo dell'Osservatorio 5, 35122 Padova, Italy}

\author{Yongzhi Cai}
\affiliation{Yunnan Observatories, Chinese Academy of Sciences, Kunming 260216, China}
\affiliation{Key Laboratory for the Structure and Evolution of Celestial Objects, Chinese Academy of Sciences, Kunming 650216, China}
\affiliation{International Centre of Supernovae, Yunnan Key Laboratory, Kunming 650216, China}

\author{Enrico Cappellaro} 
\affiliation{INAF -- Osservatorio Astronomico di Padova, Vicolo dell'Osservatorio 5, 35122 Padova, Italy}

\author{Zhihao Chen}
\affiliation{Department of Physics and Tsinghua Center for Astrophysics (THCA), Tsinghua University, Haidian District, Beijing 100084, China}

\author{Zhitong Li}
\affiliation{Key Laboratory of Space Astronomy and Technology, National Astronomical Observatories, Chinese Academy of Sciences, 20A Datun Road, Beijing 100101, China}
\affiliation{Key Laboratory of Optical Astronomy, National Astronomical Observatories, Chinese Academy of Sciences, Beijing 100101, China}

    \author{Andrea Pastorello}
\affiliation{INAF -- Osservatorio Astronomico di Padova, Vicolo dell'Osservatorio 5, 35122 Padova, Italy}

\author{Tianmeng Zhang}
\affiliation{Key Laboratory of Space Astronomy and Technology, National Astronomical Observatories, Chinese Academy of Sciences, 20A Datun Road, Beijing 100101, China}
\affiliation{School of Astronomy and Space Science, University of Chinese Academy of Sciences, Beijing 101408, China}

%% file: Exp_param_USSNe.tex
\begin{table*}[htbp]
    \label{tab: Exp_param_USSNe}
    \caption{Explosion Parameters of Suggested Ultrastripped-Envelope Supernovae$^a$}
    \centering
    
        \resizebox{\linewidth}{!}{
        \begin{tabular}{ccccccccc}
            \hline
            \hline
            Transient & Redshift & Host Galaxy & Metallicity & Peak Magnitude & Ejecta Mass            & Nickel Mass                    & Envelope Radius    & Envelope Mass \\
            Name      &          & Type        & Z$_{\odot}$&  ($r$ band)       & $M_{\rm ej}$ (\msun) & $M_{\rm Ni}$ ($10^{-2}$~\msun) & $R_{\rm ext}$ ($10^{13}$~cm) & $M_{\rm ext}$ ($10^{-2}$~\msun)\\
            
            \hline           
            SN 2021agco& 0.01056& Spiral  & 1.3 & $-16.2 \pm 0.24$ & $0.26_{-0.02}^{+0.04}$ & $2.2_{-0.3}^{+0.2}$ & $0.55_{-0.14}^{+0.18}$& $10.04_{-1.05}^{+1.87}$   \\
            
            AT 2019wxt & 0.036  & Compact &  -  & $-16.6 \pm 0.4$ & $0.20^{+0.12}_{-0.11}$ & $2.7^{+0.33}_{-0.18}$ & $35.8^{+4.06}_{-3.68}$ & $3.55^{+0.12}_{-0.11}$\\
            
            SN 2019dge & 0.021  & Compact & 0.4 & $-16.3 \pm 0.2$ & $0.30^{+0.02}_{-0.02}$ & $1.6^{+0.04}_{-0.03}$ & $1.2^{+0.06}_{-0.05}$  & $9.71^{+0.28}_{-0.27}$\\
            
            iPTF14gqr & 0.063  & Spiral  & 1.2 & $-17.5 \pm 0.2$ & $0.20^{-0.10}_{+0.10}$ & $5.0^{+0.14}_{-0.15}$   & $6.1^{+8.73}_{-3.18}$  & $2.59^{+0.46}_{-0.34}$\\
            
            iPTF16hgs & 0.017  & Spiral  & 1.4 & $-15.5 \pm 0.2$ & $1.68^{+0.28}_{-0.25}$ & $2.5^{+0.20}_{-0.22}$ & $2.6^{+14.08}_{-1.80}$ & $9.27^{+3.40}_{-2.48}$\\
            \hline
        \end{tabular}
        }
        $^a${Summary properties of the USSN candidates, including the redshift, type, and metallicity of the host galaxy, the $r$-band peak luminosity, and explosion parameters inferred from the light-curve fitting: ejecta mass $M_{\rm ej}$, nickel mass $M_{\rm Ni}$, envelope radius $R_{\rm ext}$, envelope mass $M_{\rm ext}$. The above properties are not derived by the same method, rather than refer to the values suggested by \cite{AT2019wxt_2022arxiv}, \cite{iPTF14gqr_De_2018}, and \cite{SN2019dge_2020ApJ}.} 
\end{table*}

%% file: Photomety.tex
\begin{longtable}{ccccc}
  \caption{Photometric Observations of SN 2021agco.}\label{tab:photometry}\\
\hline
MJD & Phase$^a$ & Filter & Magnitude & Instrument\\
\hline
\endfirsthead
\multicolumn{5}{c}%
{{\bfseries \tablename\ \thetable{} -- Continued}} \\
\hline 
MJD & Phase (days)$^a$\tnote{a}  & Filter & Magnitude & Instrument\\
\hline
\endhead
\hline 
\multicolumn{5}{r}{{Continued}} \\ \hline
\endfoot
\hline
\endlastfoot

59522.58 & -31.23 & $o$ & $>$19.91 & ATLAS \\
59534.51 & -19.30 & $o$ & $>$17.66 & ATLAS \\
59546.53 & -7.28 & $o$ & $>$19.77 & ATLAS \\
59550.59 & -3.22 & $o$ & 19.57$\pm$0.15 & ATLAS \\
59558.58 & +4.77 & $o$ & 18.04$\pm$0.09 & ATLAS \\
59562.49 & +8.68 & $o$ & 18.22$\pm$0.06 & ATLAS \\
59564.48 & +10.67 & $o$ & 18.59$\pm$0.14 & ATLAS \\
59566.41 & +12.60 & $o$ & 18.62$\pm$0.15 & ATLAS \\
59574.46 & +20.65 & $o$ & 19.68$\pm$0.18 & ATLAS \\
59617.44 & +63.63 & $o$ & $>$19.86 & ATLAS \\
59550.79 & -3.02 & $C$ & 18.15$\pm$0.39 & HMT \\
59552.91 & -0.90 & $C$ & 17.09$\pm$0.43 & HMT \\
59553.92 & +0.10 & $C$ & 17.31$\pm$0.40 & HMT \\
59555.77 & +1.95 & $B$ & 18.12$\pm$0.06 & TNT \\
59567.72 & +13.90 & $B$ & 19.74$\pm$0.25 & TNT \\
59582.83 & +29.02 & $B$ & 21.67$\pm$0.52 & TNT \\
59555.77 & +1.95 & $V$ & 17.82$\pm$0.04 & TNT \\
59561.77 & +7.96 & $V$ & 18.17$\pm$0.05 & TNT \\
59565.74 & +11.93 & $V$ & 19.46$\pm$0.31 & TNT \\
59567.72 & +13.90 & $V$ & 19.17$\pm$0.15 & TNT \\
59582.83 & +29.02 & $V$ & 21.30$\pm$0.26 & TNT \\
59555.77 & +1.96 & $g$ & 17.85$\pm$0.02 & TNT \\
59561.77 & +7.96 & $g$ & 18.70$\pm$0.04 & TNT \\
59567.72 & +13.91 & $g$ & 19.37$\pm$0.11 & TNT \\
59582.83 & +29.02 & $g$ & 22.22$\pm$0.38 & TNT \\
59555.77 & +1.96 & $r$ & 17.80$\pm$0.03 & TNT \\
59561.77 & +7.96 & $r$ & 18.45$\pm$0.03 & TNT \\
59565.74 & +11.93 & $r$ & 19.02$\pm$0.13 & TNT \\
59567.72 & +13.91 & $r$ & 19.14$\pm$0.09 & TNT \\
59568.66 & +14.85 & $r$ & 19.60$\pm$0.10 & TNT \\
59580.72 & +26.90 & $r$ & 20.75$\pm$0.18 & TNT \\
59582.83 & +29.02 & $r$ & 20.59$\pm$0.10 & TNT \\
59585.76 & +31.95 & $r$ & 20.93$\pm$0.15 & TNT \\
59590.63 & +36.82 & $r$ & 21.03$\pm$0.25 & TNT \\
59594.66 & +40.85 & $r$ & 20.66$\pm$0.32 & TNT \\
59596.69 & +42.87 & $r$ & 20.15$\pm$0.29 & TNT \\
59555.77 & +1.96 & $i$ & 17.80$\pm$0.03 & TNT \\
59561.77 & +7.96 & $i$ & 18.43$\pm$0.04 & TNT \\
59565.74 & +11.93 & $i$ & 19.18$\pm$0.16 & TNT \\
59582.83 & +29.02 & $i$ & 20.75$\pm$0.23 & TNT \\
59585.76 & +31.95 & $i$ & 20.72$\pm$0.20 & TNT \\
59589.64 & +35.82 & $i$ & 20.54$\pm$0.33 & TNT \\
59594.66 & +40.85 & $i$ & 20.65$\pm$0.38 & TNT \\
59553.62 & -0.20 & $B$ & 18.03$\pm$0.10 & NEXT \\
59555.64 & +1.83 & $B$ & 18.16$\pm$0.06 & NEXT \\
59556.62 & +2.80 & $B$ & 18.32$\pm$0.08 & NEXT \\
59559.56 & +5.75 & $B$ & 18.57$\pm$0.14 & NEXT \\
59561.78 & +7.97 & $B$ & 18.86$\pm$0.10 & NEXT \\
59566.70 & +12.89 & $B$ & 19.30$\pm$0.37 & NEXT \\
59570.67 & +16.86 & $B$ & 20.08$\pm$0.36 & NEXT \\
59553.62 & -0.19 & $V$ & 17.69$\pm$0.09 & NEXT \\
59555.65 & +1.83 & $V$ & 18.02$\pm$0.07 & NEXT \\
59558.59 & +4.78 & $V$ & 18.32$\pm$0.08 & NEXT \\
59559.56 & +5.75 & $V$ & 18.17$\pm$0.10 & NEXT \\
59564.63 & +10.81 & $V$ & 19.00$\pm$0.17 & NEXT \\
59565.79 & +11.98 & $V$ & 19.19$\pm$0.20 & NEXT \\
59567.61 & +13.80 & $V$ & 19.08$\pm$0.35 & NEXT \\
59570.67 & +16.86 & $V$ & 20.11$\pm$0.32 & NEXT \\
59571.72 & +17.91 & $V$ & $>$20.02 & NEXT \\
59553.62 & -0.19 & $g$ & 17.77$\pm$0.09 & NEXT \\
59555.63 & +1.82 & $g$ & 17.76$\pm$0.06 & NEXT \\
59556.63 & +2.81 & $g$ & 18.05$\pm$0.07 & NEXT \\
59558.60 & +4.78 & $g$ & 18.13$\pm$0.06 & NEXT \\
59559.57 & +5.75 & $g$ & 18.31$\pm$0.08 & NEXT \\
59564.63 & +10.82 & $g$ & 18.87$\pm$0.24 & NEXT \\
59565.80 & +11.98 & $g$ & 19.27$\pm$0.19 & NEXT \\
59570.68 & +16.86 & $g$ & $>$20.15 & NEXT \\
59571.72 & +17.91 & $g$ & $>$20.32 & NEXT \\
59553.63 & -0.19 & $r$ & 17.56$\pm$0.10 & NEXT \\
59555.63 & +1.82 & $r$ & 17.67$\pm$0.05 & NEXT \\
59556.63 & +2.82 & $r$ & 17.78$\pm$0.05 & NEXT \\
59558.60 & +4.79 & $r$ & 18.08$\pm$0.07 & NEXT \\
59559.57 & +5.76 & $r$ & 18.05$\pm$0.06 & NEXT \\
59560.61 & +6.80 & $r$ & 18.49$\pm$0.16 & NEXT \\
59564.63 & +10.82 & $r$ & 18.81$\pm$0.19 & NEXT \\
59565.80 & +11.99 & $r$ & 19.03$\pm$0.11 & NEXT \\
59570.68 & +16.87 & $r$ & 19.55$\pm$0.30 & NEXT \\
59553.63 & -0.18 & $i$ & 17.75$\pm$0.10 & NEXT \\
59555.64 & +1.82 & $i$ & 17.65$\pm$0.07 & NEXT \\
59556.63 & +2.82 & $i$ & 17.81$\pm$0.07 & NEXT \\
59558.60 & +4.79 & $i$ & 17.98$\pm$0.10 & NEXT \\
59559.57 & +5.76 & $i$ & 18.02$\pm$0.10 & NEXT \\
59564.64 & +10.82 & $i$ & 18.47$\pm$0.16 & NEXT \\
59565.80 & +11.99 & $i$ & 19.23$\pm$0.17 & NEXT \\
59567.62 & +13.81 & $i$ & 19.28$\pm$0.19 & NEXT \\
59553.64 & -0.18 & $z$ & 17.97$\pm$0.20 & NEXT \\
59555.64 & +1.83 & $z$ & 17.59$\pm$0.22 & NEXT \\
59558.61 & +4.79 & $z$ & 17.64$\pm$0.20 & NEXT \\
59559.58 & +5.76 & $z$ & 17.85$\pm$0.30 & NEXT \\
59564.64 & +10.83 & $z$ & 18.49$\pm$0.31 & NEXT \\
59565.81 & +11.99 & $z$ & 18.84$\pm$0.25 & NEXT \\
59566.72 & +12.91 & $z$ & 18.29$\pm$0.39 & NEXT \\
59570.69 & +16.87 & $z$ & 18.81$\pm$0.39 & NEXT \\
\cline{1-5}
\multicolumn{5}{c}{$^a$ Phase relative to $r$-band peak.}\\ 
\end{longtable}

%% file: SpecTable.tex
\begin{table*}
    \centering
    \caption{Observation Log of Optical Spectroscopy of SN 2021agco}
    \label{tab: spectra log}
    \begin{tabular}{ccccccc}
    \hline
    \hline
    Start Time & MJD & Phase$^a$ & Telescope+Instrument & Exposure Time & Airmass & Resolution\\%
    (UTC)      &     & (days)       &                      & (s)        &  &($\lambda$/$\Delta\lambda$)  \\
    \hline
    2022/12/06.84 & 59554.84 & 1.0 & XLT+BFOSC       & 3000 & 1.15 & 350\\ 
    2022/12/07.83 & 59555.83 & 2.0 & Copernico+AFOSC & 2700 & 1.64 & 300\\
    2022/12/11.35 & 59559.35 & 5.5 & Shane+Kast      & 2700 & 1.14 & 450\\ 
    2022/12/14.74 & 59562.74 & 8.9 & XLT+BFOSC       & 3600 & 1.04 & 350\\
    2022/12/31.70 & 59579.70 & 25.9& LJT+YFOSC       & 1800 & 1.21 & 350\\
    \hline
    \end{tabular}\\
    $^a$Phase relative to the $r$-band maximum brightness (MJD = 59553.81).
\end{table*}

%% file: bolometric.tex
\begin{table}
\caption{The evolution of bolometric light curve ($L_{\rm bb}$), temperature ($T_{\rm bb}$), and photospheric radius ($R_{\rm bb}$) of SN 2021agco, derived by fitting the SED using a blackbody function. The phases are relative to the $r$-band peak brightness.}
    \centering
    \begin{tabular}{ccccc}
\hline
\hline
MJD & Phase & $L_{\rm bb}$  & $R_{\rm bb}$  & $T_{\rm bb}$  \\
~ & (days) & ($10^{41}$~erg~s$^{-1}$) & ($10^{3}$~\rsun) &($10^{3}$~K) \\
\hline
59553.6 & -0.2 & $21.00_{-5.16}^{+10.03}$ & $4.26_{-0.81}^{+0.84}$ & $13.54_{-1.98}^{+3.01}$ \\
59555.7 & +1.9 & $13.24_{-1.60}^{+2.23}$ & $5.49_{-0.64}^{+0.68}$ & $10.62_{-0.89}^{+1.11}$ \\
59556.6 & +2.8 & $11.31_{-1.67}^{+2.60}$ & $5.79_{-0.97}^{+1.09}$ & $9.93_{-1.13}^{+1.51}$ \\
59558.6 & +4.8 & $7.71_{-1.10}^{+2.08}$ & $6.23_{-1.34}^{+1.55}$ & $8.71_{-1.19}^{+1.71}$ \\
59559.6 & +5.8 & $8.51_{-1.14}^{+1.79}$ & $5.55_{-0.96}^{+1.08}$ & $9.45_{-1.07}^{+1.42}$ \\
59561.8 & +8.0 & $7.37_{-1.20}^{+1.91}$ & $3.98_{-0.62}^{+0.68}$ & $10.77_{-1.22}^{+1.63}$ \\
59564.6 & +10.8 & $3.80_{-0.51}^{+1.07}$ & $5.57_{-1.56}^{+1.88}$ & $7.70_{-1.23}^{+1.94}$ \\
59565.8 & +12.0 & $3.88_{-0.99}^{+2.31}$ & $2.86_{-0.73}^{+0.84}$ & $10.81_{-1.95}^{+3.26}$ \\
59566.6 & +12.8 & $4.06_{-0.78}^{+1.70}$ & $5.37_{-1.87}^{+2.88}$ & $7.95_{-1.79}^{+2.75}$ \\
59567.7 & +13.9 & $3.72_{-0.78}^{+1.72}$ & $2.70_{-0.68}^{+0.77}$ & $11.01_{-1.81}^{+2.95}$ \\
59570.7 & +16.9 & $1.78_{-0.37}^{+24.37}$ & $3.68_{-3.07}^{+4.42}$ & $7.53_{-2.38}^{+30.26}$ \\
59582.8 & +29.0 & $0.45_{-0.05}^{+0.06}$ & $4.19_{-1.03}^{+1.27}$ & $5.23_{-0.54}^{+0.68}$ \\
\hline
    \end{tabular}
    \label{tab:BB_evolution}
\end{table}

%% file: EBV.tex

\begin{table*}[]
    \centering
    
    \begin{tabular}{ccccccccc}
    \hline
    \hline
    $L_{\rm peak}$ & $T_{\rm peak}$&$M_{\rm env}$ & $R_{\rm env}$ &$E_{\rm ext}$ & $E_{\rm k}$& $M_{\rm ejecta}$& $M_{\rm Ni}$  \\
    
    ($10^{42}~erg~s^{-1}$)&($10^{3}~K$) & (\msun) & (\rsun) & ($10^{49}~erg$) & ($10^{49}~erg$) & (\msun) & (\msun)  \\
    \hline
    
   $21.00_{-5.16}^{+10.03}$ & $13.54_{-1.98}^{+3.01}$ & $0.10_{-0.01}^{+0.02}$ & $78.37_{-19.94}^{+25.59}$ & $8.93_{-1.61}^{+2.59}$ & $9.57_{-1.62}^{+3.01}$ & $0.26_{-0.02}^{+0.04}$ & $0.022_{-0.003}^{+0.002}$\\
    
    \hline
    \end{tabular}
    \caption{Explosion and progenitor parameters derived from fitting to the spectral energy distribution of SN 2021agco with the shock cooling + Ni decay model.}
    \label{tab:fitresultsebv}
\end{table*}